\documentclass[12pt, twocolumn]{openjournal}
\usepackage{lipsum}
\usepackage{graphicx}
\usepackage{xcolor}
\usepackage{textgreek}
\usepackage[utf8]{inputenc}
\usepackage[english]{babel}
\usepackage{multirow}
\usepackage{hyperref}
\hypersetup{
    colorlinks=true,
    linkcolor=blue,
    filecolor=blue,      
    urlcolor=blue,
    citecolor=blue,
}
\usepackage{color,colortbl}
\usepackage{tensind}
\tensordelimiter{?}
\DeclareGraphicsExtensions{.bmp,.png,.jpg,.pdf}
\usepackage{verbatim}
\usepackage[normalem]{ulem}
\usepackage{orcidlink}
\usepackage{soul}
\graphicspath{ {./figs/} }
\def\taueff{\left\langle \tau_\mathrm{GP}\right\rangle_{50}}

\begin{document}
\title{Cosmic Reionization on Computers: Statistical Properties of the Distributions of Mean Opacities}
\author{Ella Werre\orcidlink{0009-0009-3820-4029}}
\email{werreell@msu.edu}
\affiliation{NSF Research Experience for Undergraduates; University of Michigan}
\affiliation{Department of Physics; Michigan State University}

\author{David Robinson\orcidlink{0000-0002-3751-6145}}
\email{dbrobins@umich.edu}
\affiliation{Department of Physics; University of Michigan, Ann Arbor, MI 48109, USA}
\affiliation{Leinweber Center for Theoretical Physics; University of Michigan, Ann Arbor, MI 48109, USA}

\author{Camille Avestruz\orcidlink{0000-0001-8868-0810}}
\affiliation{Department of Physics; University of Michigan, Ann Arbor, MI 48109, USA}
\affiliation{Leinweber Center for Theoretical Physics; University of Michigan, Ann Arbor, MI 48109, USA}

\author{Nickolay Y. Gnedin\orcidlink{0000-0001-5925-4580}}
\affiliation{Theory Division; 
Fermi National Accelerator Laboratory;
Batavia, IL 60510, USA}
\affiliation{Kavli Institute for Cosmological Physics;
The University of Chicago;
Chicago, IL 60637, USA}
\affiliation{Department of Astronomy \& Astrophysics; 
The University of Chicago; 
Chicago, IL 60637, USA}

\begin{abstract}
Quasar absorption lines provide a unique window to the relationship between galaxies and the intergalactic medium during the Epoch of Reionization.  In particular, high redshift quasars enable measurements of the neutral hydrogen content of the universe.  However, the limited sample size of observed quasar spectra, particularly at the highest redshifts, hampers our ability to fully characterize the intergalactic medium during this epoch from observations alone. In this work, we characterize the distributions of mean opacities of the intergalactic medium in simulations from the Cosmic Reionization on Computers (CROC) project.  We find that the distribution of mean opacities along sightlines follows a non-trivial distribution that cannot be easily approximated by a known distribution.  When comparing the cumulative distribution function of mean opacities measurements in subsamples of sample sizes similar to observational measurements from the literature, we find consistency between CROC and observations at redshifts $z\lesssim 5.7$.  However, at higher redshifts ($z\gtrsim5.7$), the cumulative distribution function of mean opacities from CROC is notably narrower than those from observed quasar sightlines implying that observations probe a systematically more opaque intergalactic medium at higher redshifts than the intergalactic medium in CROC boxes at these same redshifts.  This is consistent with previous analyses that indicate that the universe is reionized too early in CROC simulations.
\end{abstract}


\maketitle

\section{Introduction}
\label{sec:intro}

During the Epoch of Reionization, ultraviolet radiation emitted by early quasars and galaxies ionized almost all of the hydrogen in the intergalactic medium \citep[IGM,][]{loebandfurlanetto2013, Eilers2018}. Prior to this epoch, most of the hydrogen in the universe had been neutral ever since the early universe cooled enough for electrons and protons to combine to form neutral hydrogen atoms \citep{dicke65}. 

The reionization process is known to have been spatially inhomogeneous \citep[e.g.][]{bahcall65, arons70, arons72}. Reionization sources create bubbles of ionized hydrogen. As the Epoch of Reionization progresses, these bubbles grow in size, eventually overlapping to fill the entire IGM \citep[e.g.][]{arons70, arons72}. The end of the Epoch of Reionization is only weakly constrained by current observational data \citep[see the reviews of][]{Becker2015, fan23}.

Sources called quasars, which are powered by accretion onto supermassive black holes at the center of high redshift galaxies, are luminous enough to be observed at high redshifts corresponding to the Epoch of Reionization \citep[see][for a recent review]{fan23}. We can study the IGM during reionization using the Lyman-alpha (Ly$\alpha$) forest, a collection of absorption lines in redshifted quasar spectra caused by resonant scattering of Lyman-alpha photons by neutral hydrogen.

The Ly$\alpha$ forest arises from interactions between neutral hydrogen clouds and photons emitted by quasars, and traces reionization along isolated sightlines through the IGM \citep{Becker2001}. Light from quasars is redshifted on its journey to our telescopes. Any neutral hydrogen clouds along the sightline between the telescope and the quasar absorb photons at the Ly$\alpha$ wavelength (1215 \r{A}) \textit{in the absorber's restframe} \citep{Laursen2011}. So, the absorption lines of the Ly$\alpha$ forest appear \textit{blueward} of the Ly$\alpha$ emission from a quasar. The neutral hydrogen fraction (and hence absorption in the Ly$\alpha$ forest) decreases as reionization progresses. When quasar light passes through regions of the IGM with high neutral hydrogenopacity, nearly all the quasar emission in a wavelength band blueward of Ly$\alpha$ will be absorbed, resulting in the Gunn-Peterson trough \citep{Gunn1965}. This trough has been detected in a quasar at $z=6.28$ \citep{Becker2001}.

Observed high redshift quasar spectra exhibit Ly$\alpha$ forest absorption lines, enabling measurements of the neutral hydrogen content at different redshifts along the quasar sightline \citep[e.g.][]{Becker2001}. However, quasars are relatively rare, making it difficult to obtain a large number of samples that probe the high redshift IGM. The decrease in available observed quasars with increasing redshift makes the interpretation of IGM measurements at high redshift even more challenging \citep[e.g.][]{Becker2015b, Nasir2020, Bosman2020sample}. One of the more recent studies into these high-redshift quasars is \citet{Bosman2022}, which examined 67 quasar sightlines to quantify the distribution of observed Ly$\alpha$ transmission measurements. This study supports other observational studies that suggest a later end of reionization, around $z\gtrsim 5$ \citep[e.g.][]{Becker2015b, Kulkarni2019}.

Simulations of the early universe act as a necessary complement to observational research. We can conduct numerical experiments to compare simulations against observational measurements and help interpret those measurements. In particular, the limited number of observed quasars available to sample the IGM at high redshifts merits a comparison with a wide range of simulations using varied methodologies.

The THESAN simulation suite models the IGM and the structures responsible for hydrogen ionization in order to further identify and characterize these sources. Different simulations in the suite assume that high ($M > 10^{10} M_\odot$) or low ($M < 10^{10} M_\odot$) mass galaxies dominate the reionization process. In the low-mass simulation, reionization is complete by $z=6$, while the high-mass simulation retains significant amounts of neutral hydrogen in the IGM until the simulation ends at $z=5.5$ \citep{Kannan_2021}. \citet{Kulkarni2019} used the Sherwood simulation suite to estimate a later end to reionization at $z \gtrsim 5.3$. The ATON-HE simulation suite incorporates the midpoint of the reionization history as an input parameter, allowing for comparisons between observations and different reionization history models \citep{Asthana_2024}. 

In this work, we use the Cosmic Reionization on Computers (CROC) simulations, which self-consistently incorporate many of the physical processes relevant during the Epoch of Reionization, and includes realizations with different reionization histories \citep{Gnedin2014}. We explore the variation and evolution of the neutral hydrogen content in CROC by quantifying the distribution of mean opacities along random sightlines from the simulations and comparing it with observational measurements.

We structure the paper as follows. Section 2 describes the CROC simulation suite and the sightlines that we analyze in this work.  We describe our analysis of the simulation data in Section 3. Section 4 presents our comparison between the mean opacity distributions from CROC data and observational data from \citet{Bosman2018} and \citet{Bosman2022}. Section 5 includes a summary of our conclusions and a discussion of the implications of our results.

\section{Data: Sightlines from the CROC Simulation Suite}
\label{sec:data}

The Cosmic Reionization on Computers (CROC) simulations model the process of cosmic reionization of hydrogen. CROC includes several simulation boxes, each with different reionization histories.  We use the boxes labeled `A', `C', and `F', which have distinct reionization start times and durations.  They are, respectively, boxes with ``median", ``early", and ``late" reionization histories. They experience a midpoint of neutral fraction at $z_\mathrm{A} \approx 8$, $z_\mathrm{C} \approx 8.2$, and $z_\mathrm{F} \approx 7.4$ respectively \citep{Gnedin2014}. From these boxes, we extract simulated sightlines with a length of $50h^{-1}$ comoving Mpc, from which we can calculate the mean optical depth of gas along each sightline.  We extract a total of 1000 sightlines from each box at every saved snapshot.  This allows us to mimic quasar sightlines that probe the evolution and gas content across varied environments of the high redshift universe.
We post-process the extracted sightlines to mimic observational data. Specifically, we mimic Lyman~$\alpha$ forest absorption lines due to neutral hydrogen in the intergalactic medium along a sightline. 

We quantify the Ly$\alpha$ forest opacity with the `effective optical depth', a quantity easily measured in observations of quasar spectra. Here, $\tau = -\ln\left<F/F_0\right>$, where $F$ is the transmitted flux, $F_0$ is the flux emitted by the quasar, and averaging is done over a spectral region corresponding to the comoving distance of $50h^{-1}$ Mpc. Since other averaging distances have been used in the literature as well, we label this optical depth measured in our simulations as $\langle \tau_{\mathrm{GP}} \rangle_{50}$ to explicitly refer to the averaging distance used. A higher optical depth indicates that the IGM contains more neutral hydrogen, thus optical depth is expected to correlate with the density of neutral hydrogen in the IGM. 

From these sightlines, we extract $\langle \tau_\mathrm{GP} \rangle_{50}$ to evaluate the scatter of mean opacity values present in the IGM during reionization \citep{Gnedin2017}. We split the optical depth data from various snapshots into redshift bins for direct comparison with observational studies. For consistent comparison to observations, we utilize sightlines from snapshots of CROC boxes in the redshift range $5.1 < z < 6.1$.

\begin{figure}[!h]
    \centering
    \includegraphics[width=0.46\textwidth]{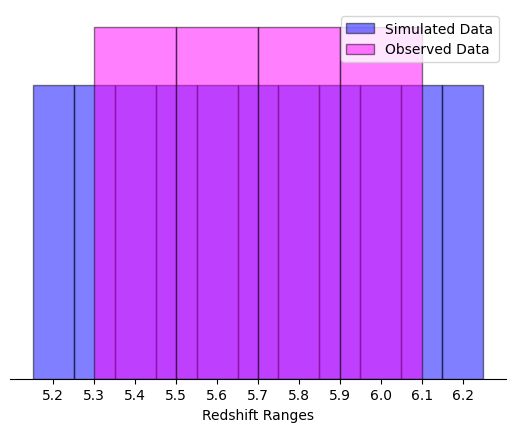}\hfill
    \caption{A cartoon diagram illustrating the overlap in the simulation snapshots (blue) and observed redshift bins (pink).  The lowest observed redshift bin ($5.3<z<5.5$) overlaps with half of the second simulation snapshot, all of the third, and half of the fourth.  We therefore construct simulated sightline samples with the same proportion of sightlines from those snapshots to match the redshift range of observed data, i.e. 1500:3000:1500 simulation sightlines to compare against observed data in the lowest redshift bin.}
    \label{fig:redshift_bins}
\end{figure}

We illustrate our approach to sampling across simulation snapshots in Fig.~\ref{fig:redshift_bins}.  The blue short histogram bin centers correspond to the discrete snapshots at which we have saved CROC box data. For each snapshot, we have 3000 simulated sightlines (1000 from each of the 3 CROC boxes we use). 

The taller pink histogram edges correspond to the bin edges for the redshift ranges considered in observations of \citet{Bosman2018} and \citet{Bosman2022}.  For example, as we sample simulated sightlines for the first bin of observed data ($5.3<z<5.5$), we see that the observed bin overlaps with half of the second blue bin, all of the third blue bin, and half of the fourth blue bin.  We therefore create a simulated data set by drawing 1500 lines of sight randomly from our second snapshot, all 3000 from the third, and 1500 from the fourth for a total of 6000 lines of sight in our simulated sample to compare against the $5.3<z<5.5$ observed data.  

\section{Methods: Quantification and Comparisons of Mean Opacities}
\label{sec:methods}

\subsection{Characterizing Mean Opacity Distributions}
\label{methods:tau_quantiles}

We quantify the distribution of mean opacities with the cumulative distribution function (CDF), with mean opacities measured in $50h^{-1}$ comoving Mpc sightlines, $\left\langle \tau_\mathrm{GP}\right\rangle_{50}$, across all three CROC boxes considered and across simulation snapshots consistent with observed redshift bins.  

\begin{figure}[!h]
    \centering\includegraphics[width=0.46\textwidth]{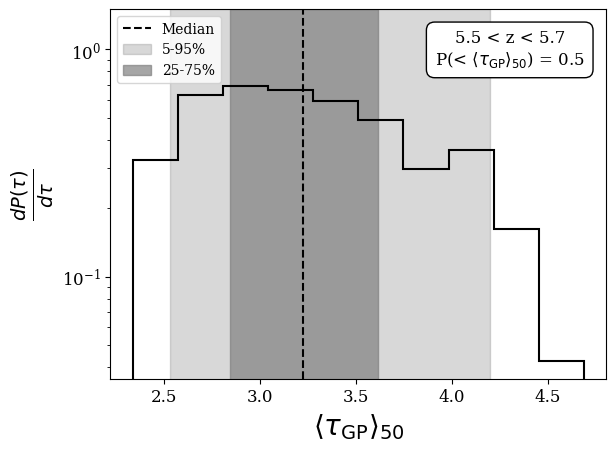}\hfill
    \caption{The distribution of mean opacities at the CDF = 0.5 point from all subsamples of 10 sightlines for the simulated sample representing the $5.7 < z < 5.9$ redshift bin.  The distribution cannot be approximated by a Gaussian, and we indicate the median with a black dashed line, the 25-75th percentile with a dark gray band, and the 5-95th percentile with the light gray band.}
    \label{fig:histogram}
\end{figure}

To capture the scatter across different sightline subsamples in the simulated sample we draw 1000 subsamples of a given size from the 6000 sightlines.  For each subsample, we construct a CDF of mean opacities in that subsample.  We next find the $\left\langle \tau_\mathrm{GP}\right\rangle_{50}$ values corresponding to a fixed point in the CDF for all 1000 subsamples and create a histogram of those values to quantify the distribution across the subsamples.  As an example, Fig.~\ref{fig:histogram} shows the distribution of $\left\langle \tau_\mathrm{GP}\right\rangle_{50}$ for subsamples of size 10 at the CDF value of 50\%, or where $P(<\left\langle \tau_\mathrm{GP}\right\rangle_{50}) = 0.5$.  Note, that this distribution cannot be described by a Gaussian, nor is it log-normal.  We therefore use quantiles to quantify the shape of this distribution. The dashed vertical line corresponds to the median value of the histogram, the dark gray band indicates the 25-75th percentile range, and the light gray band is the 5-95th percentile range.  For reference, we also include histograms at CDF values of 10\% and 90\% in the Appendix, Fig.~\ref{fig:appendix_histograms}.  These also cannot be described by a known distribution.  And, for completeness, the Appendix Fig.~\ref{fig:croc-pdf-tau} also includes the full CDFs at each redshift bin, where each CDF value includes percentile bands corresponding to the distribution from all 1000 subsamples of size 10 drawn from the simulated sightlines.

\subsection{Constructing Statistically Consistent Comparisons with Observations}
\label{methods:obs}

To make statistically consistent comparisons between the mean opacity distributions in CROC simulations and observational data, we compare the shape and width of the CDF described in Section~\ref{methods:tau_quantiles} with subsamples of the same size as those in observations.  For example, the number of quasar sightlines in the lowest redshift bin of \citet{Bosman2018} is 57.  To compare against these specific measurements, we, therefore, create subsamples of size 57 instead of subsamples of size 10 that were used to construct Fig.~\ref{fig:histogram} and the distributions in the Appendix figures.  Hereafter, each comparison of CROC simulations vs the observational data uses subsamples with the appropriately matched sizes.  For reference, we summarize these in the Appendix Table~\ref{tab:redshift_objects}.

Note, if we were to simply take sightlines from the CROC snapshot at the midpoint of the redshift range corresponding to an observed subsample, our sightlines would not be a fair representation of that sample.  We show example CDFs from biased sightline selections in the Appendix Fig.~\ref{fig:croc-pdf-fixed-z-comparison}.  During these epochs, the mean opacity distribution evolves over the redshift ranges selected in observed data.  The CDF of CROC sightlines drawn from only the midpoint epoch has systematically more sightlines at larger mean opacity than the CDF of CROC sightlines drawn from a sample across the entire redshift range that matches the observational redshift bin.

\section{Results}
\label{sec:results}

\subsection{Comparisons with Observations}
\label{res:obs_cdfs}

If we are to use simulations to characterize the statistical properties of the distribution of effective opacities in the forest, the simulations must be reasonably realistic. To check that, we compare mean opacity measurements from simulated sightlines in CROC to quasar observations from \citet{Bosman2018} and \citet{Bosman2022}.  The goal of this section is to determine how well the simulation reproduces the observed distributions of Lyman-$\alpha$ mean opacities. 

\begin{table*}[ht]
 \centering
 \renewcommand{\arraystretch}{1.5} 
 \setlength{\tabcolsep}{10pt} 
 \caption{Observed sightline count in each redshift bin from \citet{Bosman2018} and \citet{Bosman2022}, which determines the subsample size in our construction of CROC CDFs and percentile spread (see Section~\ref{methods:tau_quantiles}). We also include the fraction of sightlines whose mean opacity measurements are upper limits. }
 \label{tab:redshift_objects}
 \begin{tabular}{|c|c|c|c|c|}
 \hline
 \multirow{2}{*}{Redshift Range} & \multicolumn{2}{c|}{Number of Objects} & \multicolumn{2}{c|}{Fraction with Upper Limit\footnote{some data only has upper limit on flux, corresponding to lower limit on mean opacity}} \\ \cline{2-5}
 & \citet{Bosman2018} & \citet{Bosman2022} & \citet{Bosman2018} & \citet{Bosman2022}\\[0.5ex]
 \hline
 $5.3 < z < 5.5$ & $57$ & $64$ & $1/57$ & $0$\\ 
 \hline
 $5.5 < z < 5.7$ & $51$ & $60$ & $5/51$ & $2/60$ \\
 \hline
 $5.7 < z < 5.9$ & $33$ & $45$ & $11/33$ & $7/45$ \\
 \hline
 $5.9 < z < 6.1$ & $11$ & $17$ & $4/11$ & $5/17$\\
 \hline
\end{tabular}
\end{table*}

Our first comparison is with \citet{Bosman2018}, which utilizes 62 quasar sightlines, with 3 previously unrecorded examples.  Our second comparison is with \citet{Bosman2022}, which utilizes 67 quasars.   Not all quasar sightlines exhibit absorption at the redshift range of interest, the number of sightlines that contribute to each redshift range are listed in Table~\ref{tab:redshift_objects}.

To account for the uncertainty in certain values of the measured flux, the observational analysis introduced ``optimistic" and ``pessimistic" bounds. The optimistic bounds are defined as the average intrinsic flux equaling twice the average flux error, while the pessimistic bound represents ``maximum opaqueness", with a value of $\tau$ approaching infinity \citep{Bosman2018}. 

\begin{figure*}[!h]
    \centering
    \includegraphics[width=0.5\textwidth]{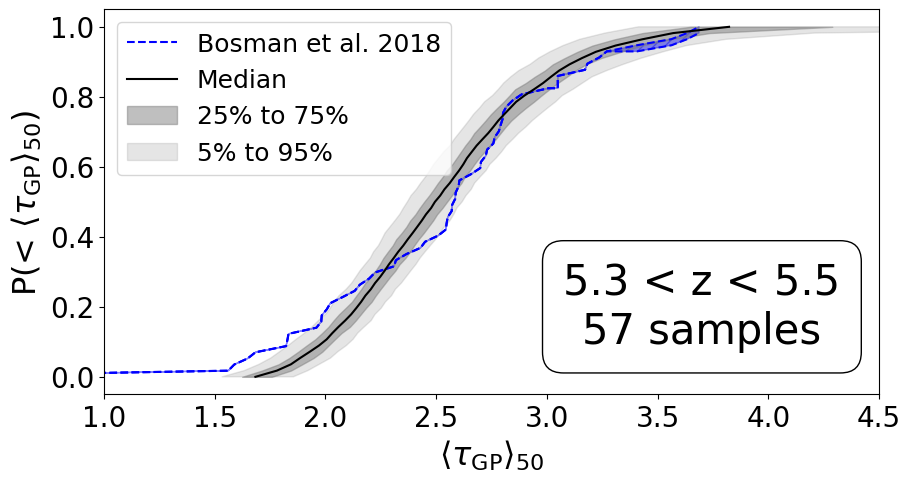}\hfill
    \includegraphics[width=0.5\textwidth]{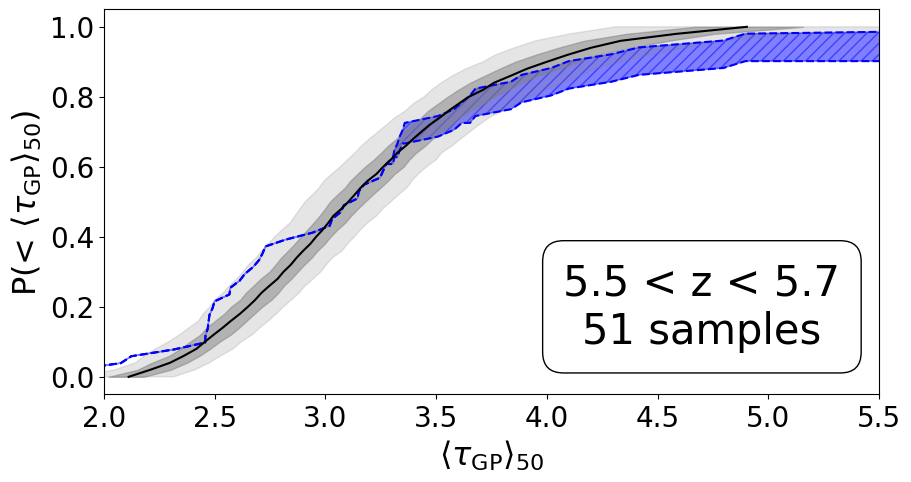}\\[1ex] 
    \includegraphics[width=0.5\textwidth]{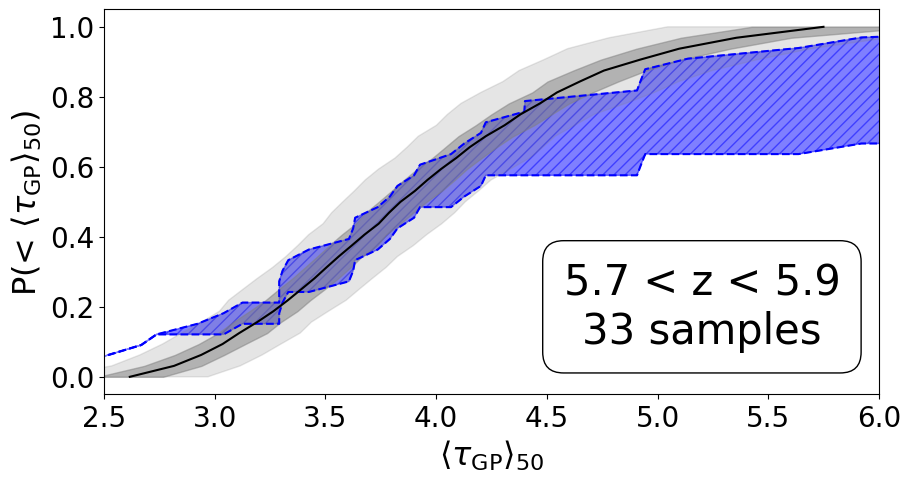}\hfill
    \includegraphics[width=0.5\textwidth]{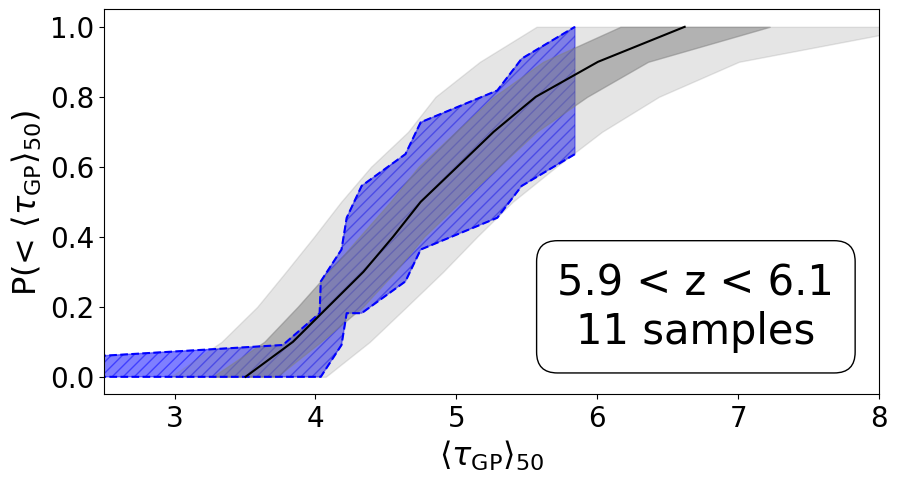}
    \caption{Comparisons between the CDF of mean opacities in CROC with those from \citet{Bosman2018} observational data. The optical depth distributions from CROC best match the observed distributions from this dataset at $z<5.7$.  At higher redshifts, the distributions still overlap within the 5th-95th percentile bands of the CROC data.} 
    \label{fig:Bos2018}
\end{figure*}

In Fig.~\ref{fig:Bos2018}, we overlay the CDF of mean opacities from CROC with the resulting CDF from the observations presented in \citet{Bosman2018}.  Each panel shows the comparison within the annotated redshift range corresponding to the labeled sample size.  Note, the sample size of observations determines the size of the subsamples we draw when constructing the CDF of mean opacities in CROC.   

With a similar color-coding and linestyle as Fig.~\ref{fig:histogram}, we outline the median value of the CDF of mean opacities from CROC with the solid black line.  At each value of the CDF, we calculate the 25th-75th percentile range and 5th-95th percentile range to determine the respective widths of the dark gray and light gray shaded regions. The shaded blue CDF corresponds to measurements in \citet{Bosman2018}.  

In the lowest two redshift bins ($z<5.7$, top two panels of Fig.~\ref{fig:Bos2018}, the CDF from the observations is consistent with the CDF from CROC with overlaps in the 25th-75th percentile region of the CROC distribution.  In the third redshift bin ($5.7<z<5.9$), the CDF from CROC exhibits a steeper distribution than the CDF from \citet{Bosman2018}, but the overlap at the high CDF values is still within the 5th-95th percentile region of the CROC distribution.  In the highest redshift bin ($5.9<z<6.1$), both CDFs well-overlap.  Note, however, that the sample size of 11 leads to fairly wide CDFs.  

\begin{figure*}[!h]
    \centering
    \includegraphics[width=0.5\textwidth]{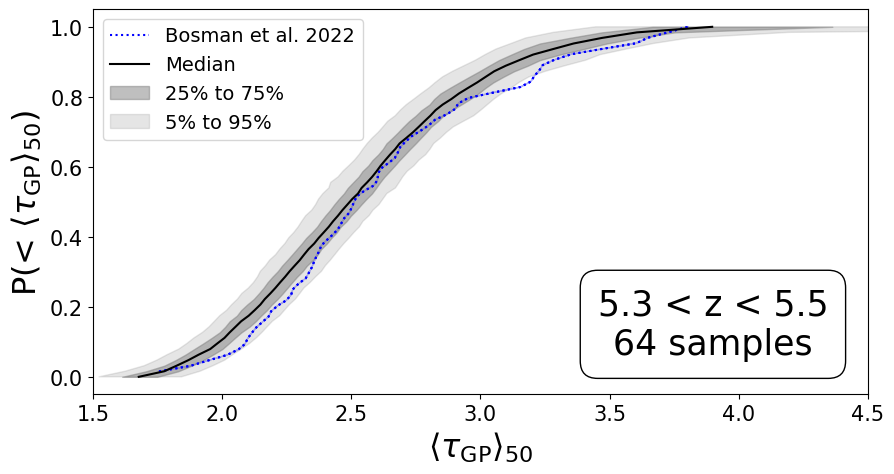}\hfill
    \includegraphics[width=0.5\textwidth]{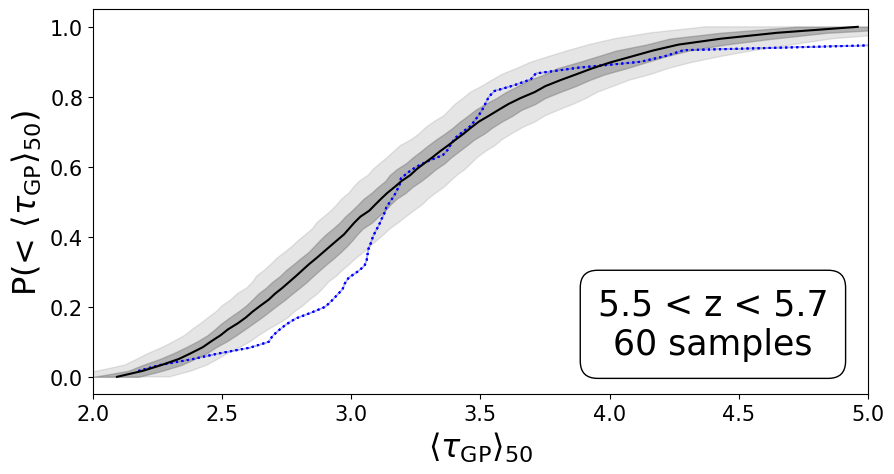}\\[1ex] 
    \includegraphics[width=0.5\textwidth]{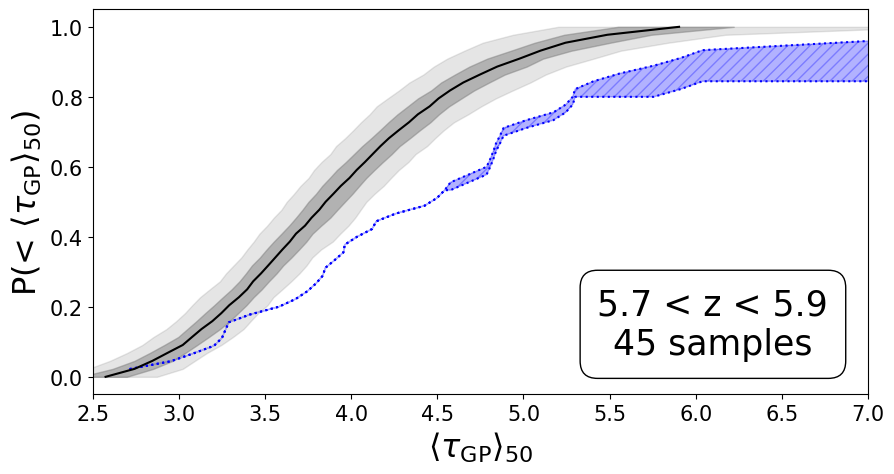}\hfill
    \includegraphics[width=0.5\textwidth]{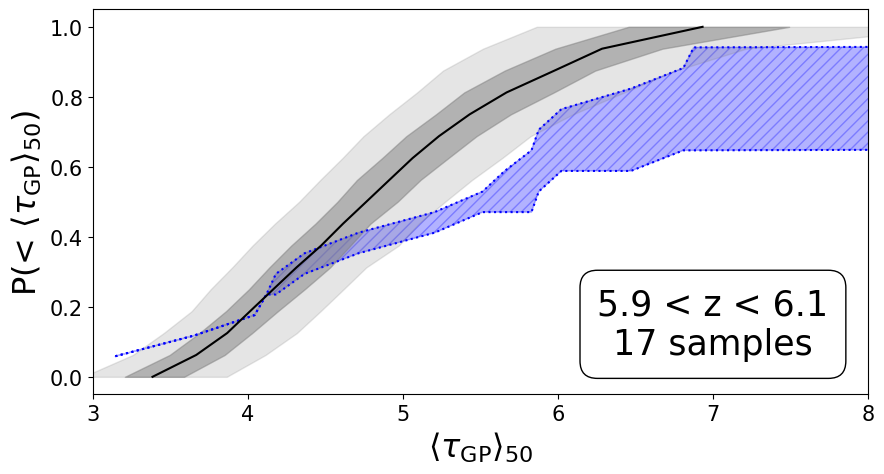}
    \caption{Comparisons between the CDF of mean opacities in CROC with those from \citet{Bosman2022} observational data. The distribution of optical depths in CROC is consistent with the CDF in observations in the lowest two redshift ranges but is notably steeper at the highest two redshift ranges.}
    \label{fig:Bos2022}
\end{figure*}

In Fig.~\ref{fig:Bos2022}, we make the same comparison to the mean opacity measurements from \citet{Bosman2022}. Note, the observational data from \citet{Bosman2022} includes more quasars, leading to a larger sample size in each redshift bin.  The larger sample sizes provide tighter constraints on simulations.

Similar to the comparison in Fig.~\ref{fig:Bos2018}, the CDF from the observations in the lowest redshift bin is statistically consistent with the distribution of mean opacities in CROC sightlines.  In the second redshift bin, there appears to be a slight deviation in the shape of the CDF, potentially indicating more inhomogeneity and relative patchiness in the IGM probed by the observed quasar sightlines in this redshift range. 

Notably, there is the most discrepancy in the two highest redshift bins. The CDF from CROC measurements is steeper than that from observations, with the observed CDF lying outside of the 5th-95th percentile of the CROC distributions at the larger CDF values.  This discrepancy (and a similar smaller discrepancy for the \citet{Bosman2018} data set) is not particularly surprising, as the CROC simulations are known to have cosmic reionization complete too early \citep{Gnedin2022}. 

However, we highlight the stark difference in mean opacities at $5.9<z<6.1$ between Fig.~\ref{fig:Bos2018} and Fig.~\ref{fig:Bos2022}.  With only 11 samples, the distribution of mean opacities from observed sightlines is completely consistent with those from our simulations.  Then, with only 6 additional observed sightlines contributing to the observational constraint of mean opacity distributions, we are able to identify a statistically significant difference between the CDF from the observations and the CDF constructed from CROC sightlines. The shallower CDF from observed sightlines hints at a more opaque IGM at these epochs.  We characterize the impact of the number of sightlines in the next subsection and further discuss the implications of the difference in the shape of the CDF in Sec.~\ref{sec:discussion}.

\subsection{Dependence of the CDF on Number of Sightlines}
\label{res:cdf_width}
\begin{figure}[!h]
    \centering
    \includegraphics[width=0.46\textwidth]{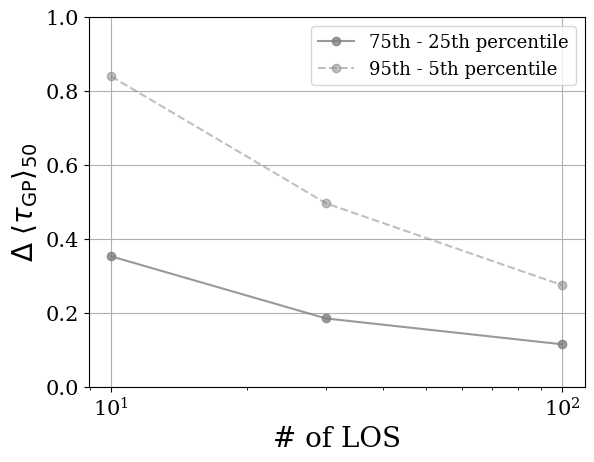}\\[1ex] 
    \caption{Width of the CDF at a fixed value of $P(<\left\langle \tau_\mathrm{GP}\right\rangle_{50})=0.5$ for the $5.5<z<5.7$ bin as a function of number of sightlines. We show both the width of the 75th-25th percentile and the 95th-5th percentile.}
    \label{fig:Varlos}
\end{figure}

The width of the CDF distribution varies with $\taueff$ and the value of the CDF. Physically, it is more informative to quantify the width of this distribution "horizontally", as a range in $\taueff$. As a representative width, we choose the range of $\taueff$ at a value for the CDF of $P(<\taueff)=0.5$, the same as shown in Fig.\ \ref{fig:histogram}. Fig.~\ref{fig:Varlos} shows the widths of the CDF of mean opacities from simulated sightlines as a function of the number of sightlines per subsample used in the redshift range of $5.5<z<5.7$.   

The dark and light gray data points for $10$ sightlines respectively correspond to the widths of the dark and light gray vertical regions in the histogram shown in Fig.~\ref{fig:histogram}.  The width of the CDF decreases as the number of sightlines $N$ increases approximately as $N^{-1/2}$, as can be expected from a simple error-on-the-mean scaling. 

For reference, we include the same tests for fixed CDF values of $P(<\left\langle \tau_\mathrm{GP}\right\rangle_{50})=0.1$ and $P(<\left\langle \tau_\mathrm{GP}\right\rangle_{50})=0.9$ for multiple redshift ranges in Fig.~\ref{fig:Varlos_appendix} in the Appendix. The widths of the CDF at fixed value of $P(<\left\langle \tau_\mathrm{GP}\right\rangle_{50})=0.1$ are near identical to those at 0.5 with similar tightening over the order of magnitude increase in number of sightlines.  However, the widths of the CDF at fixed value of $P(<\left\langle \tau_\mathrm{GP}\right\rangle_{50})=0.9$ are everywhere larger, although the scaling with $N$ remains the same.

\section{Conclusions and Discussion}\label{sec:discussion}

In this work, we quantify the distributions of the mean Gunn-Peterson opacities of the intergalactic medium measured in random sightlines from the Cosmic Reionization on Computers simulation suite.  We quantify the cumulative distribution function (CDF) of Ly$\alpha$ mean opacities measured from simulated sightlines with a length of $50h^{-1}$ comoving Mpc.  We calculate widths of the CDF at any given fixed point by constructing a histogram of of CDF values by subsampling our sightlines with a chosen subsample size.  Varying the subsample size allows us to make statistically consistent comparisons with observational data from \citet{Bosman2018} and \citet{Bosman2022}.  We further construct simulated samples of sightlines with a redshift distribution consistent with those in the observational samples we compare against.  We additionally quantify the dependence of CDF width on the number of sightlines in the subsampling procedure.  Our summary of our conclusions are as follows. 

\begin{itemize}
    \item The distribution of mean opacities at a fixed CDF value is not well-described by a Gaussian or log-normal (see Fig.~\ref{fig:histogram}).
    \item The distributions of mean opacities in CROC simulations are consistent with observational data from \citet{Bosman2018} and \citet{Bosman2022} for $z<5.7$ but become discrepant at higher redshifts (except for the $5.9<z<6.1$ redshift bin in \citet{Bosman2018}, which is itself significantly discrepant from the same redshift bin in \citet{Bosman2022}).    
    \item The width of the simulated mean opacity distribution decreases with the number of sightlines $N$ approximately as $N^{-1/2}$ (see Fig.~\ref{fig:Varlos}).  
    \item The disagreement between observations and CROC at $z\gtrsim5.7$ implies that the observed sightlines probe a systematically more opaque IGM than the IGM in CROC at those same redshifts.  This finding indicates that our universe was more opaque than CROC at the same redshifts.  We note that our finding is consistent with the fact that all individual CROC boxes have higher reionization redshifts ($z>6.5$) compared to those inferred from observations, which indicate a potential reionization time below $z\lesssim6$ \citep[e.g.][]{Becker2015b, Eilers2018, keating20, Nasir2020, zhu21, becker21, Bosman2022, zhu22, zhu23}.  Previous comparisons between the mean free path in CROC simulations and observations also indicate that the universe is reionized too early in CROC simulations \citet{fanMFPCROC2024}.
\end{itemize}

Finally, we comment on potential biases that might affect our comparison between the mean opacity distribution in CROC and the distribution from observations. First, we randomly draw sightlines in our simulation.  This, by construction, probes the entire intergalactic medium in our simulation boxes.  True quasar sightlines, however, start at the centers of massive halos, which biases the sightline origin to overdense regions. Observational studies however explicitly remove spectral regions in the vicinity of quasars to avoid such bias, so our comparison with the observational data is expected to be fair and unbiased.

\section*{Acknowledgments}
EW thanks the National Science Foundation and the University of Michigan Department of Physics for support during the Summer 2024 Research Experience for Undergraduates Program. DR and CA thank the Leinweber Center for Theoretical Physics for support during this project. This manuscript has also been co-authored by Fermi Research Alliance, LLC under Contract No. DE-AC02-07CH11359 with the United States Department of Energy. This work used resources of the Argonne Leadership Computing Facility, which is a DOE Office of Science User Facility supported under Contract DE-AC02-06CH11357. An award of computer time was provided by the Innovative and Novel Computational Impact on Theory and Experiment (INCITE) program. This research is also part of the Blue Waters sustained-petascale computing project, which is supported by the National Science Foundation (awards OCI-0725070 and ACI-1238993) and the state of Illinois. Blue Waters is a joint effort of the University of Illinois at Urbana-Champaign and its National Center for Supercomputing Applications. The repository that contains the code for this analysis can be found at:  \url{https://github.com/werreell/Simulated-Quasar-Sightline-Analysis-with-CROC}.

\bibliographystyle{apsrev4-1}
\bibliography{references}

\begin{thebibliography}{28}%
\makeatletter
\providecommand \@ifxundefined [1]{%
 \@ifx{#1\undefined}
}%
\providecommand \@ifnum [1]{%
 \ifnum #1\expandafter \@firstoftwo
 \else \expandafter \@secondoftwo
 \fi
}%
\providecommand \@ifx [1]{%
 \ifx #1\expandafter \@firstoftwo
 \else \expandafter \@secondoftwo
 \fi
}%
\providecommand \natexlab [1]{#1}%
\providecommand \enquote  [1]{``#1''}%
\providecommand \bibnamefont  [1]{#1}%
\providecommand \bibfnamefont [1]{#1}%
\providecommand \citenamefont [1]{#1}%
\providecommand \href@noop [0]{\@secondoftwo}%
\providecommand \href [0]{\begingroup \@sanitize@url \@href}%
\providecommand \@href[1]{\@@startlink{#1}\@@href}%
\providecommand \@@href[1]{\endgroup#1\@@endlink}%
\providecommand \@sanitize@url [0]{\catcode `\\12\catcode `\$12\catcode `\&12\catcode `\#12\catcode `\^12\catcode `\_12\catcode `\%12\relax}%
\providecommand \@@startlink[1]{}%
\providecommand \@@endlink[0]{}%
\providecommand \url  [0]{\begingroup\@sanitize@url \@url }%
\providecommand \@url [1]{\endgroup\@href {#1}{\urlprefix }}%
\providecommand \urlprefix  [0]{URL }%
\providecommand \Eprint [0]{\href }%
\providecommand \doibase [0]{http://dx.doi.org/}%
\providecommand \selectlanguage [0]{\@gobble}%
\providecommand \bibinfo  [0]{\@secondoftwo}%
\providecommand \bibfield  [0]{\@secondoftwo}%
\providecommand \translation [1]{[#1]}%
\providecommand \BibitemOpen [0]{}%
\providecommand \bibitemStop [0]{}%
\providecommand \bibitemNoStop [0]{.\EOS\space}%
\providecommand \EOS [0]{\spacefactor3000\relax}%
\providecommand \BibitemShut  [1]{\csname bibitem#1\endcsname}%
\let\auto@bib@innerbib\@empty
\bibitem [{\citenamefont {{Loeb}}\ and\ \citenamefont {{Furlanetto}}(2013)}]{loebandfurlanetto2013}%
  \BibitemOpen
  \bibfield  {author} {\bibinfo {author} {\bibfnamefont {A.}~\bibnamefont {{Loeb}}}\ and\ \bibinfo {author} {\bibfnamefont {S.~R.}\ \bibnamefont {{Furlanetto}}},\ }\href@noop {} {\emph {\bibinfo {title} {{The First Galaxies in the Universe}}}}\ (\bibinfo {year} {2013})\BibitemShut {NoStop}%
\bibitem [{\citenamefont {{Eilers}}\ \emph {et~al.}(2018)\citenamefont {{Eilers}}, \citenamefont {{Davies}},\ and\ \citenamefont {{Hennawi}}}]{Eilers2018}%
  \BibitemOpen
  \bibfield  {author} {\bibinfo {author} {\bibfnamefont {A.-C.}\ \bibnamefont {{Eilers}}}, \bibinfo {author} {\bibfnamefont {F.~B.}\ \bibnamefont {{Davies}}}, \ and\ \bibinfo {author} {\bibfnamefont {J.~F.}\ \bibnamefont {{Hennawi}}},\ }\href {\doibase 10.3847/1538-4357/aad4fd} {\bibfield  {journal} {\bibinfo  {journal} {\apj}\ }\textbf {\bibinfo {volume} {864}},\ \bibinfo {eid} {53} (\bibinfo {year} {2018})},\ \Eprint {http://arxiv.org/abs/1807.04229} {arXiv:1807.04229 [astro-ph.GA]} \BibitemShut {NoStop}%
\bibitem [{\citenamefont {{Dicke}}\ \emph {et~al.}(1965)\citenamefont {{Dicke}}, \citenamefont {{Peebles}}, \citenamefont {{Roll}},\ and\ \citenamefont {{Wilkinson}}}]{dicke65}%
  \BibitemOpen
  \bibfield  {author} {\bibinfo {author} {\bibfnamefont {R.~H.}\ \bibnamefont {{Dicke}}}, \bibinfo {author} {\bibfnamefont {P.~J.~E.}\ \bibnamefont {{Peebles}}}, \bibinfo {author} {\bibfnamefont {P.~G.}\ \bibnamefont {{Roll}}}, \ and\ \bibinfo {author} {\bibfnamefont {D.~T.}\ \bibnamefont {{Wilkinson}}},\ }\href {\doibase 10.1086/148306} {\bibfield  {journal} {\bibinfo  {journal} {\apj}\ }\textbf {\bibinfo {volume} {142}},\ \bibinfo {pages} {414} (\bibinfo {year} {1965})}\BibitemShut {NoStop}%
\bibitem [{\citenamefont {{Bahcall}}\ and\ \citenamefont {{Salpeter}}(1965)}]{bahcall65}%
  \BibitemOpen
  \bibfield  {author} {\bibinfo {author} {\bibfnamefont {J.~N.}\ \bibnamefont {{Bahcall}}}\ and\ \bibinfo {author} {\bibfnamefont {E.~E.}\ \bibnamefont {{Salpeter}}},\ }\href {\doibase 10.1086/148460} {\bibfield  {journal} {\bibinfo  {journal} {\apj}\ }\textbf {\bibinfo {volume} {142}},\ \bibinfo {pages} {1677} (\bibinfo {year} {1965})}\BibitemShut {NoStop}%
\bibitem [{\citenamefont {{Arons}}\ and\ \citenamefont {{McCray}}(1970)}]{arons70}%
  \BibitemOpen
  \bibfield  {author} {\bibinfo {author} {\bibfnamefont {J.}~\bibnamefont {{Arons}}}\ and\ \bibinfo {author} {\bibfnamefont {R.}~\bibnamefont {{McCray}}},\ }\href@noop {} {\bibfield  {journal} {\bibinfo  {journal} {\aplett}\ }\textbf {\bibinfo {volume} {5}},\ \bibinfo {pages} {123} (\bibinfo {year} {1970})}\BibitemShut {NoStop}%
\bibitem [{\citenamefont {{Arons}}\ and\ \citenamefont {{Wingert}}(1972)}]{arons72}%
  \BibitemOpen
  \bibfield  {author} {\bibinfo {author} {\bibfnamefont {J.}~\bibnamefont {{Arons}}}\ and\ \bibinfo {author} {\bibfnamefont {D.~W.}\ \bibnamefont {{Wingert}}},\ }\href {\doibase 10.1086/151682} {\bibfield  {journal} {\bibinfo  {journal} {\apj}\ }\textbf {\bibinfo {volume} {177}},\ \bibinfo {pages} {1} (\bibinfo {year} {1972})}\BibitemShut {NoStop}%
\bibitem [{\citenamefont {{Becker}}\ \emph {et~al.}(2015)\citenamefont {{Becker}}, \citenamefont {{Bolton}},\ and\ \citenamefont {{Lidz}}}]{Becker2015}%
  \BibitemOpen
  \bibfield  {author} {\bibinfo {author} {\bibfnamefont {G.~D.}\ \bibnamefont {{Becker}}}, \bibinfo {author} {\bibfnamefont {J.~S.}\ \bibnamefont {{Bolton}}}, \ and\ \bibinfo {author} {\bibfnamefont {A.}~\bibnamefont {{Lidz}}},\ }\href {\doibase 10.1017/pasa.2015.45} {\bibfield  {journal} {\bibinfo  {journal} {\pasa}\ }\textbf {\bibinfo {volume} {32}},\ \bibinfo {eid} {e045} (\bibinfo {year} {2015})},\ \Eprint {http://arxiv.org/abs/1510.03368} {arXiv:1510.03368 [astro-ph.CO]} \BibitemShut {NoStop}%
\bibitem [{\citenamefont {Fan}\ \emph {et~al.}(2023)\citenamefont {Fan}, \citenamefont {Bañados},\ and\ \citenamefont {Simcoe}}]{fan23}%
  \BibitemOpen
  \bibfield  {author} {\bibinfo {author} {\bibfnamefont {X.}~\bibnamefont {Fan}}, \bibinfo {author} {\bibfnamefont {E.}~\bibnamefont {Bañados}}, \ and\ \bibinfo {author} {\bibfnamefont {R.~A.}\ \bibnamefont {Simcoe}},\ }\href {\doibase https://doi.org/10.1146/annurev-astro-052920-102455} {\bibfield  {journal} {\bibinfo  {journal} {Annual Review of Astronomy and Astrophysics}\ }\textbf {\bibinfo {volume} {61}},\ \bibinfo {pages} {373} (\bibinfo {year} {2023})}\BibitemShut {NoStop}%
\bibitem [{\citenamefont {{Becker}}\ \emph {et~al.}(2001)\citenamefont {{Becker}}, \citenamefont {{Fan}}, \citenamefont {{White}}, \citenamefont {{Strauss}}, \citenamefont {{Narayanan}}, \citenamefont {{Lupton}}, \citenamefont {{Gunn}}, \citenamefont {{Annis}}, \citenamefont {{Bahcall}}, \citenamefont {{Brinkmann}}, \citenamefont {{Connolly}}, \citenamefont {{Csabai}}, \citenamefont {{Czarapata}}, \citenamefont {{Doi}}, \citenamefont {{Heckman}}, \citenamefont {{Hennessy}}, \citenamefont {{Ivezi{\'c}}}, \citenamefont {{Knapp}}, \citenamefont {{Lamb}}, \citenamefont {{McKay}}, \citenamefont {{Munn}}, \citenamefont {{Nash}}, \citenamefont {{Nichol}}, \citenamefont {{Pier}}, \citenamefont {{Richards}}, \citenamefont {{Schneider}}, \citenamefont {{Stoughton}}, \citenamefont {{Szalay}}, \citenamefont {{Thakar}},\ and\ \citenamefont {{York}}}]{Becker2001}%
  \BibitemOpen
  \bibfield  {author} {\bibinfo {author} {\bibfnamefont {R.~H.}\ \bibnamefont {{Becker}}}, \bibinfo {author} {\bibfnamefont {X.}~\bibnamefont {{Fan}}}, \bibinfo {author} {\bibfnamefont {R.~L.}\ \bibnamefont {{White}}}, \bibinfo {author} {\bibfnamefont {M.~A.}\ \bibnamefont {{Strauss}}}, \bibinfo {author} {\bibfnamefont {V.~K.}\ \bibnamefont {{Narayanan}}}, \bibinfo {author} {\bibfnamefont {R.~H.}\ \bibnamefont {{Lupton}}}, \bibinfo {author} {\bibfnamefont {J.~E.}\ \bibnamefont {{Gunn}}}, \bibinfo {author} {\bibfnamefont {J.}~\bibnamefont {{Annis}}}, \bibinfo {author} {\bibfnamefont {N.~A.}\ \bibnamefont {{Bahcall}}}, \bibinfo {author} {\bibfnamefont {J.}~\bibnamefont {{Brinkmann}}}, \bibinfo {author} {\bibfnamefont {A.~J.}\ \bibnamefont {{Connolly}}}, \bibinfo {author} {\bibfnamefont {I.}~\bibnamefont {{Csabai}}}, \bibinfo {author} {\bibfnamefont {P.~C.}\ \bibnamefont {{Czarapata}}}, \bibinfo {author} {\bibfnamefont {M.}~\bibnamefont {{Doi}}}, \bibinfo {author} {\bibfnamefont {T.~M.}\ \bibnamefont
  {{Heckman}}}, \bibinfo {author} {\bibfnamefont {G.~S.}\ \bibnamefont {{Hennessy}}}, \bibinfo {author} {\bibfnamefont {{\v{Z}}.}~\bibnamefont {{Ivezi{\'c}}}}, \bibinfo {author} {\bibfnamefont {G.~R.}\ \bibnamefont {{Knapp}}}, \bibinfo {author} {\bibfnamefont {D.~Q.}\ \bibnamefont {{Lamb}}}, \bibinfo {author} {\bibfnamefont {T.~A.}\ \bibnamefont {{McKay}}}, \bibinfo {author} {\bibfnamefont {J.~A.}\ \bibnamefont {{Munn}}}, \bibinfo {author} {\bibfnamefont {T.}~\bibnamefont {{Nash}}}, \bibinfo {author} {\bibfnamefont {R.}~\bibnamefont {{Nichol}}}, \bibinfo {author} {\bibfnamefont {J.~R.}\ \bibnamefont {{Pier}}}, \bibinfo {author} {\bibfnamefont {G.~T.}\ \bibnamefont {{Richards}}}, \bibinfo {author} {\bibfnamefont {D.~P.}\ \bibnamefont {{Schneider}}}, \bibinfo {author} {\bibfnamefont {C.}~\bibnamefont {{Stoughton}}}, \bibinfo {author} {\bibfnamefont {A.~S.}\ \bibnamefont {{Szalay}}}, \bibinfo {author} {\bibfnamefont {A.~R.}\ \bibnamefont {{Thakar}}}, \ and\ \bibinfo {author} {\bibfnamefont {D.~G.}\ \bibnamefont
  {{York}}},\ }\href {\doibase 10.1086/324231} {\bibfield  {journal} {\bibinfo  {journal} {\aj}\ }\textbf {\bibinfo {volume} {122}},\ \bibinfo {pages} {2850} (\bibinfo {year} {2001})},\ \Eprint {http://arxiv.org/abs/astro-ph/0108097} {arXiv:astro-ph/0108097 [astro-ph]} \BibitemShut {NoStop}%
\bibitem [{\citenamefont {Laursen}\ \emph {et~al.}(2011)\citenamefont {Laursen}, \citenamefont {Sommer-Larsen},\ and\ \citenamefont {Razoumov}}]{Laursen2011}%
  \BibitemOpen
  \bibfield  {author} {\bibinfo {author} {\bibfnamefont {P.}~\bibnamefont {Laursen}}, \bibinfo {author} {\bibfnamefont {J.}~\bibnamefont {Sommer-Larsen}}, \ and\ \bibinfo {author} {\bibfnamefont {A.}~\bibnamefont {Razoumov}},\ }\href {\doibase 10.1088/0004-637X/728/1/52} {\bibfield  {journal} {\bibinfo  {journal} {The Astrophysical Journal}\ }\textbf {\bibinfo {volume} {728}} (\bibinfo {year} {2011}),\ 10.1088/0004-637X/728/1/52}\BibitemShut {NoStop}%
\bibitem [{\citenamefont {{Gunn}}\ and\ \citenamefont {{Peterson}}(1965)}]{Gunn1965}%
  \BibitemOpen
  \bibfield  {author} {\bibinfo {author} {\bibfnamefont {J.~E.}\ \bibnamefont {{Gunn}}}\ and\ \bibinfo {author} {\bibfnamefont {B.~A.}\ \bibnamefont {{Peterson}}},\ }\href {\doibase 10.1086/148444} {\bibfield  {journal} {\bibinfo  {journal} {\apj}\ }\textbf {\bibinfo {volume} {142}},\ \bibinfo {pages} {1633} (\bibinfo {year} {1965})}\BibitemShut {NoStop}%
\bibitem [{\citenamefont {Becker}\ \emph {et~al.}(2015)\citenamefont {Becker}, \citenamefont {Bolton}, \citenamefont {Madau}, \citenamefont {Pettini}, \citenamefont {Ryan-Weber},\ and\ \citenamefont {Venemans}}]{Becker2015b}%
  \BibitemOpen
  \bibfield  {author} {\bibinfo {author} {\bibfnamefont {G.~D.}\ \bibnamefont {Becker}}, \bibinfo {author} {\bibfnamefont {J.~S.}\ \bibnamefont {Bolton}}, \bibinfo {author} {\bibfnamefont {P.}~\bibnamefont {Madau}}, \bibinfo {author} {\bibfnamefont {M.}~\bibnamefont {Pettini}}, \bibinfo {author} {\bibfnamefont {E.~V.}\ \bibnamefont {Ryan-Weber}}, \ and\ \bibinfo {author} {\bibfnamefont {B.~P.}\ \bibnamefont {Venemans}},\ }\href {\doibase 10.1093/mnras/stu2646} {\bibfield  {journal} {\bibinfo  {journal} {Monthly Notices of the Royal Astronomical Society}\ }\textbf {\bibinfo {volume} {447}},\ \bibinfo {pages} {3402} (\bibinfo {year} {2015})},\ \Eprint {http://arxiv.org/abs/https://academic.oup.com/mnras/article-pdf/447/4/3402/5695265/stu2646.pdf} {https://academic.oup.com/mnras/article-pdf/447/4/3402/5695265/stu2646.pdf} \BibitemShut {NoStop}%
\bibitem [{\citenamefont {Nasir}\ and\ \citenamefont {D’Aloisio}(2020)}]{Nasir2020}%
  \BibitemOpen
  \bibfield  {author} {\bibinfo {author} {\bibfnamefont {F.}~\bibnamefont {Nasir}}\ and\ \bibinfo {author} {\bibfnamefont {A.}~\bibnamefont {D’Aloisio}},\ }\href {\doibase 10.1093/mnras/staa894} {\bibfield  {journal} {\bibinfo  {journal} {Monthly Notices of the Royal Astronomical Society}\ }\textbf {\bibinfo {volume} {494}},\ \bibinfo {pages} {3080–3094} (\bibinfo {year} {2020})}\BibitemShut {NoStop}%
\bibitem [{\citenamefont {{Bosman}}(2020)}]{Bosman2020sample}%
  \BibitemOpen
  \bibfield  {author} {\bibinfo {author} {\bibfnamefont {S.}~\bibnamefont {{Bosman}}},\ }\href {\doibase 10.5281/zenodo.3634964} {\enquote {\bibinfo {title} {{All z$>$5.7 quasars currently known}},}\ }\bibinfo {howpublished} {Zenodo dataset} (\bibinfo {year} {2020})\BibitemShut {NoStop}%
\bibitem [{\citenamefont {{Bosman}}\ \emph {et~al.}(2022)\citenamefont {{Bosman}}, \citenamefont {{Davies}}, \citenamefont {{Becker}}, \citenamefont {{Keating}}, \citenamefont {{Davies}}, \citenamefont {{Zhu}}, \citenamefont {{Eilers}}, \citenamefont {{D'Odorico}}, \citenamefont {{Bian}}, \citenamefont {{Bischetti}}, \citenamefont {{Cristiani}}, \citenamefont {{Fan}}, \citenamefont {{Farina}}, \citenamefont {{Haehnelt}}, \citenamefont {{Hennawi}}, \citenamefont {{Kulkarni}}, \citenamefont {{Mesinger}}, \citenamefont {{Meyer}}, \citenamefont {{Onoue}}, \citenamefont {{Pallottini}}, \citenamefont {{Qin}}, \citenamefont {{Ryan-Weber}}, \citenamefont {{Schindler}}, \citenamefont {{Walter}}, \citenamefont {{Wang}},\ and\ \citenamefont {{Yang}}}]{Bosman2022}%
  \BibitemOpen
  \bibfield  {author} {\bibinfo {author} {\bibfnamefont {S.~E.~I.}\ \bibnamefont {{Bosman}}}, \bibinfo {author} {\bibfnamefont {F.~B.}\ \bibnamefont {{Davies}}}, \bibinfo {author} {\bibfnamefont {G.~D.}\ \bibnamefont {{Becker}}}, \bibinfo {author} {\bibfnamefont {L.~C.}\ \bibnamefont {{Keating}}}, \bibinfo {author} {\bibfnamefont {R.~L.}\ \bibnamefont {{Davies}}}, \bibinfo {author} {\bibfnamefont {Y.}~\bibnamefont {{Zhu}}}, \bibinfo {author} {\bibfnamefont {A.-C.}\ \bibnamefont {{Eilers}}}, \bibinfo {author} {\bibfnamefont {V.}~\bibnamefont {{D'Odorico}}}, \bibinfo {author} {\bibfnamefont {F.}~\bibnamefont {{Bian}}}, \bibinfo {author} {\bibfnamefont {M.}~\bibnamefont {{Bischetti}}}, \bibinfo {author} {\bibfnamefont {S.~V.}\ \bibnamefont {{Cristiani}}}, \bibinfo {author} {\bibfnamefont {X.}~\bibnamefont {{Fan}}}, \bibinfo {author} {\bibfnamefont {E.~P.}\ \bibnamefont {{Farina}}}, \bibinfo {author} {\bibfnamefont {M.~G.}\ \bibnamefont {{Haehnelt}}}, \bibinfo {author} {\bibfnamefont {J.~F.}\ \bibnamefont
  {{Hennawi}}}, \bibinfo {author} {\bibfnamefont {G.}~\bibnamefont {{Kulkarni}}}, \bibinfo {author} {\bibfnamefont {A.}~\bibnamefont {{Mesinger}}}, \bibinfo {author} {\bibfnamefont {R.~A.}\ \bibnamefont {{Meyer}}}, \bibinfo {author} {\bibfnamefont {M.}~\bibnamefont {{Onoue}}}, \bibinfo {author} {\bibfnamefont {A.}~\bibnamefont {{Pallottini}}}, \bibinfo {author} {\bibfnamefont {Y.}~\bibnamefont {{Qin}}}, \bibinfo {author} {\bibfnamefont {E.}~\bibnamefont {{Ryan-Weber}}}, \bibinfo {author} {\bibfnamefont {J.-T.}\ \bibnamefont {{Schindler}}}, \bibinfo {author} {\bibfnamefont {F.}~\bibnamefont {{Walter}}}, \bibinfo {author} {\bibfnamefont {F.}~\bibnamefont {{Wang}}}, \ and\ \bibinfo {author} {\bibfnamefont {J.}~\bibnamefont {{Yang}}},\ }\href {\doibase 10.1093/mnras/stac1046} {\bibfield  {journal} {\bibinfo  {journal} {\mnras}\ }\textbf {\bibinfo {volume} {514}},\ \bibinfo {pages} {55} (\bibinfo {year} {2022})},\ \Eprint {http://arxiv.org/abs/2108.03699} {arXiv:2108.03699 [astro-ph.CO]} \BibitemShut {NoStop}%
\bibitem [{\citenamefont {{Kulkarni}}\ \emph {et~al.}(2019)\citenamefont {{Kulkarni}}, \citenamefont {{Keating}}, \citenamefont {{Haehnelt}}, \citenamefont {{Bosman}}, \citenamefont {{Puchwein}}, \citenamefont {{Chardin}},\ and\ \citenamefont {{Aubert}}}]{Kulkarni2019}%
  \BibitemOpen
  \bibfield  {author} {\bibinfo {author} {\bibfnamefont {G.}~\bibnamefont {{Kulkarni}}}, \bibinfo {author} {\bibfnamefont {L.~C.}\ \bibnamefont {{Keating}}}, \bibinfo {author} {\bibfnamefont {M.~G.}\ \bibnamefont {{Haehnelt}}}, \bibinfo {author} {\bibfnamefont {S.~E.~I.}\ \bibnamefont {{Bosman}}}, \bibinfo {author} {\bibfnamefont {E.}~\bibnamefont {{Puchwein}}}, \bibinfo {author} {\bibfnamefont {J.}~\bibnamefont {{Chardin}}}, \ and\ \bibinfo {author} {\bibfnamefont {D.}~\bibnamefont {{Aubert}}},\ }\href {\doibase 10.1093/mnrasl/slz025} {\bibfield  {journal} {\bibinfo  {journal} {\mnras}\ }\textbf {\bibinfo {volume} {485}},\ \bibinfo {pages} {L24} (\bibinfo {year} {2019})},\ \Eprint {http://arxiv.org/abs/1809.06374} {arXiv:1809.06374 [astro-ph.CO]} \BibitemShut {NoStop}%
\bibitem [{\citenamefont {Kannan}\ \emph {et~al.}(2021)\citenamefont {Kannan}, \citenamefont {Garaldi}, \citenamefont {Smith}, \citenamefont {Pakmor}, \citenamefont {Springel}, \citenamefont {Vogelsberger},\ and\ \citenamefont {Hernquist}}]{Kannan_2021}%
  \BibitemOpen
  \bibfield  {author} {\bibinfo {author} {\bibfnamefont {R.}~\bibnamefont {Kannan}}, \bibinfo {author} {\bibfnamefont {E.}~\bibnamefont {Garaldi}}, \bibinfo {author} {\bibfnamefont {A.}~\bibnamefont {Smith}}, \bibinfo {author} {\bibfnamefont {R.}~\bibnamefont {Pakmor}}, \bibinfo {author} {\bibfnamefont {V.}~\bibnamefont {Springel}}, \bibinfo {author} {\bibfnamefont {M.}~\bibnamefont {Vogelsberger}}, \ and\ \bibinfo {author} {\bibfnamefont {L.}~\bibnamefont {Hernquist}},\ }\href {\doibase 10.1093/mnras/stab3710} {\bibfield  {journal} {\bibinfo  {journal} {Monthly Notices of the Royal Astronomical Society}\ }\textbf {\bibinfo {volume} {511}},\ \bibinfo {pages} {4005–4030} (\bibinfo {year} {2021})}\BibitemShut {NoStop}%
\bibitem [{\citenamefont {Asthana}\ \emph {et~al.}(2024)\citenamefont {Asthana}, \citenamefont {Haehnelt}, \citenamefont {Kulkarni}, \citenamefont {Aubert}, \citenamefont {Bolton},\ and\ \citenamefont {Keating}}]{Asthana_2024}%
  \BibitemOpen
  \bibfield  {author} {\bibinfo {author} {\bibfnamefont {S.}~\bibnamefont {Asthana}}, \bibinfo {author} {\bibfnamefont {M.~G.}\ \bibnamefont {Haehnelt}}, \bibinfo {author} {\bibfnamefont {G.}~\bibnamefont {Kulkarni}}, \bibinfo {author} {\bibfnamefont {D.}~\bibnamefont {Aubert}}, \bibinfo {author} {\bibfnamefont {J.~S.}\ \bibnamefont {Bolton}}, \ and\ \bibinfo {author} {\bibfnamefont {L.~C.}\ \bibnamefont {Keating}},\ }\href {\doibase 10.1093/mnras/stae1945} {\bibfield  {journal} {\bibinfo  {journal} {Monthly Notices of the Royal Astronomical Society}\ }\textbf {\bibinfo {volume} {533}},\ \bibinfo {pages} {2843–2866} (\bibinfo {year} {2024})}\BibitemShut {NoStop}%
\bibitem [{\citenamefont {{Gnedin}}(2014)}]{Gnedin2014}%
  \BibitemOpen
  \bibfield  {author} {\bibinfo {author} {\bibfnamefont {N.~Y.}\ \bibnamefont {{Gnedin}}},\ }\href {\doibase 10.1088/0004-637X/793/1/29} {\bibfield  {journal} {\bibinfo  {journal} {\apj}\ }\textbf {\bibinfo {volume} {793}},\ \bibinfo {eid} {29} (\bibinfo {year} {2014})},\ \Eprint {http://arxiv.org/abs/1403.4245} {arXiv:1403.4245 [astro-ph.CO]} \BibitemShut {NoStop}%
\bibitem [{\citenamefont {{Bosman}}\ \emph {et~al.}(2018)\citenamefont {{Bosman}}, \citenamefont {{Fan}}, \citenamefont {{Jiang}}, \citenamefont {{Reed}}, \citenamefont {{Matsuoka}}, \citenamefont {{Becker}},\ and\ \citenamefont {{Haehnelt}}}]{Bosman2018}%
  \BibitemOpen
  \bibfield  {author} {\bibinfo {author} {\bibfnamefont {S.~E.~I.}\ \bibnamefont {{Bosman}}}, \bibinfo {author} {\bibfnamefont {X.}~\bibnamefont {{Fan}}}, \bibinfo {author} {\bibfnamefont {L.}~\bibnamefont {{Jiang}}}, \bibinfo {author} {\bibfnamefont {S.}~\bibnamefont {{Reed}}}, \bibinfo {author} {\bibfnamefont {Y.}~\bibnamefont {{Matsuoka}}}, \bibinfo {author} {\bibfnamefont {G.}~\bibnamefont {{Becker}}}, \ and\ \bibinfo {author} {\bibfnamefont {M.}~\bibnamefont {{Haehnelt}}},\ }\href {\doibase 10.1093/mnras/sty1344} {\bibfield  {journal} {\bibinfo  {journal} {\mnras}\ }\textbf {\bibinfo {volume} {479}},\ \bibinfo {pages} {1055} (\bibinfo {year} {2018})},\ \Eprint {http://arxiv.org/abs/1802.08177} {arXiv:1802.08177 [astro-ph.GA]} \BibitemShut {NoStop}%
\bibitem [{\citenamefont {{Gnedin}}\ \emph {et~al.}(2017)\citenamefont {{Gnedin}}, \citenamefont {{Becker}},\ and\ \citenamefont {{Fan}}}]{Gnedin2017}%
  \BibitemOpen
  \bibfield  {author} {\bibinfo {author} {\bibfnamefont {N.~Y.}\ \bibnamefont {{Gnedin}}}, \bibinfo {author} {\bibfnamefont {G.~D.}\ \bibnamefont {{Becker}}}, \ and\ \bibinfo {author} {\bibfnamefont {X.}~\bibnamefont {{Fan}}},\ }\href {\doibase 10.3847/1538-4357/aa6c24} {\bibfield  {journal} {\bibinfo  {journal} {\apj}\ }\textbf {\bibinfo {volume} {841}},\ \bibinfo {eid} {26} (\bibinfo {year} {2017})},\ \Eprint {http://arxiv.org/abs/1605.03183} {arXiv:1605.03183 [astro-ph.CO]} \BibitemShut {NoStop}%
\bibitem [{\citenamefont {{Gnedin}}(2022)}]{Gnedin2022}%
  \BibitemOpen
  \bibfield  {author} {\bibinfo {author} {\bibfnamefont {N.~Y.}\ \bibnamefont {{Gnedin}}},\ }\href {\doibase 10.3847/1538-4357/ac8d0a} {\bibfield  {journal} {\bibinfo  {journal} {\apj}\ }\textbf {\bibinfo {volume} {937}},\ \bibinfo {eid} {17} (\bibinfo {year} {2022})},\ \Eprint {http://arxiv.org/abs/2204.05338} {arXiv:2204.05338 [astro-ph.CO]} \BibitemShut {NoStop}%
\bibitem [{\citenamefont {{Keating}}\ \emph {et~al.}(2020)\citenamefont {{Keating}}, \citenamefont {{Weinberger}}, \citenamefont {{Kulkarni}}, \citenamefont {{Haehnelt}}, \citenamefont {{Chardin}},\ and\ \citenamefont {{Aubert}}}]{keating20}%
  \BibitemOpen
  \bibfield  {author} {\bibinfo {author} {\bibfnamefont {L.~C.}\ \bibnamefont {{Keating}}}, \bibinfo {author} {\bibfnamefont {L.~H.}\ \bibnamefont {{Weinberger}}}, \bibinfo {author} {\bibfnamefont {G.}~\bibnamefont {{Kulkarni}}}, \bibinfo {author} {\bibfnamefont {M.~G.}\ \bibnamefont {{Haehnelt}}}, \bibinfo {author} {\bibfnamefont {J.}~\bibnamefont {{Chardin}}}, \ and\ \bibinfo {author} {\bibfnamefont {D.}~\bibnamefont {{Aubert}}},\ }\href {\doibase 10.1093/mnras/stz3083} {\bibfield  {journal} {\bibinfo  {journal} {\mnras}\ }\textbf {\bibinfo {volume} {491}},\ \bibinfo {pages} {1736} (\bibinfo {year} {2020})},\ \Eprint {http://arxiv.org/abs/1905.12640} {arXiv:1905.12640 [astro-ph.CO]} \BibitemShut {NoStop}%
\bibitem [{\citenamefont {{Zhu}}\ \emph {et~al.}(2021)\citenamefont {{Zhu}}, \citenamefont {{Becker}}, \citenamefont {{Bosman}}, \citenamefont {{Keating}}, \citenamefont {{Christenson}}, \citenamefont {{Ba{\~n}ados}}, \citenamefont {{Bian}}, \citenamefont {{Davies}}, \citenamefont {{D'Odorico}}, \citenamefont {{Eilers}}, \citenamefont {{Fan}}, \citenamefont {{Haehnelt}}, \citenamefont {{Kulkarni}}, \citenamefont {{Pallottini}}, \citenamefont {{Qin}}, \citenamefont {{Wang}},\ and\ \citenamefont {{Yang}}}]{zhu21}%
  \BibitemOpen
  \bibfield  {author} {\bibinfo {author} {\bibfnamefont {Y.}~\bibnamefont {{Zhu}}}, \bibinfo {author} {\bibfnamefont {G.~D.}\ \bibnamefont {{Becker}}}, \bibinfo {author} {\bibfnamefont {S.~E.~I.}\ \bibnamefont {{Bosman}}}, \bibinfo {author} {\bibfnamefont {L.~C.}\ \bibnamefont {{Keating}}}, \bibinfo {author} {\bibfnamefont {H.~M.}\ \bibnamefont {{Christenson}}}, \bibinfo {author} {\bibfnamefont {E.}~\bibnamefont {{Ba{\~n}ados}}}, \bibinfo {author} {\bibfnamefont {F.}~\bibnamefont {{Bian}}}, \bibinfo {author} {\bibfnamefont {F.~B.}\ \bibnamefont {{Davies}}}, \bibinfo {author} {\bibfnamefont {V.}~\bibnamefont {{D'Odorico}}}, \bibinfo {author} {\bibfnamefont {A.-C.}\ \bibnamefont {{Eilers}}}, \bibinfo {author} {\bibfnamefont {X.}~\bibnamefont {{Fan}}}, \bibinfo {author} {\bibfnamefont {M.~G.}\ \bibnamefont {{Haehnelt}}}, \bibinfo {author} {\bibfnamefont {G.}~\bibnamefont {{Kulkarni}}}, \bibinfo {author} {\bibfnamefont {A.}~\bibnamefont {{Pallottini}}}, \bibinfo {author} {\bibfnamefont {Y.}~\bibnamefont {{Qin}}},
  \bibinfo {author} {\bibfnamefont {F.}~\bibnamefont {{Wang}}}, \ and\ \bibinfo {author} {\bibfnamefont {J.}~\bibnamefont {{Yang}}},\ }\href {\doibase 10.3847/1538-4357/ac26c2} {\bibfield  {journal} {\bibinfo  {journal} {\apj}\ }\textbf {\bibinfo {volume} {923}},\ \bibinfo {eid} {223} (\bibinfo {year} {2021})},\ \Eprint {http://arxiv.org/abs/2109.06295} {arXiv:2109.06295 [astro-ph.CO]} \BibitemShut {NoStop}%
\bibitem [{\citenamefont {{Becker}}\ \emph {et~al.}(2021)\citenamefont {{Becker}}, \citenamefont {{D'Aloisio}}, \citenamefont {{Christenson}}, \citenamefont {{Zhu}}, \citenamefont {{Worseck}},\ and\ \citenamefont {{Bolton}}}]{becker21}%
  \BibitemOpen
  \bibfield  {author} {\bibinfo {author} {\bibfnamefont {G.~D.}\ \bibnamefont {{Becker}}}, \bibinfo {author} {\bibfnamefont {A.}~\bibnamefont {{D'Aloisio}}}, \bibinfo {author} {\bibfnamefont {H.~M.}\ \bibnamefont {{Christenson}}}, \bibinfo {author} {\bibfnamefont {Y.}~\bibnamefont {{Zhu}}}, \bibinfo {author} {\bibfnamefont {G.}~\bibnamefont {{Worseck}}}, \ and\ \bibinfo {author} {\bibfnamefont {J.~S.}\ \bibnamefont {{Bolton}}},\ }\href {\doibase 10.1093/mnras/stab2696} {\bibfield  {journal} {\bibinfo  {journal} {\mnras}\ }\textbf {\bibinfo {volume} {508}},\ \bibinfo {pages} {1853} (\bibinfo {year} {2021})},\ \Eprint {http://arxiv.org/abs/2103.16610} {arXiv:2103.16610 [astro-ph.CO]} \BibitemShut {NoStop}%
\bibitem [{\citenamefont {{Zhu}}\ \emph {et~al.}(2022)\citenamefont {{Zhu}}, \citenamefont {{Becker}}, \citenamefont {{Bosman}}, \citenamefont {{Keating}}, \citenamefont {{D'Odorico}}, \citenamefont {{Davies}}, \citenamefont {{Christenson}}, \citenamefont {{Ba{\~n}ados}}, \citenamefont {{Bian}}, \citenamefont {{Bischetti}}, \citenamefont {{Chen}}, \citenamefont {{Davies}}, \citenamefont {{Eilers}}, \citenamefont {{Fan}}, \citenamefont {{Gaikwad}}, \citenamefont {{Greig}}, \citenamefont {{Haehnelt}}, \citenamefont {{Kulkarni}}, \citenamefont {{Lai}}, \citenamefont {{Pallottini}}, \citenamefont {{Qin}}, \citenamefont {{Ryan-Weber}}, \citenamefont {{Walter}}, \citenamefont {{Wang}},\ and\ \citenamefont {{Yang}}}]{zhu22}%
  \BibitemOpen
  \bibfield  {author} {\bibinfo {author} {\bibfnamefont {Y.}~\bibnamefont {{Zhu}}}, \bibinfo {author} {\bibfnamefont {G.~D.}\ \bibnamefont {{Becker}}}, \bibinfo {author} {\bibfnamefont {S.~E.~I.}\ \bibnamefont {{Bosman}}}, \bibinfo {author} {\bibfnamefont {L.~C.}\ \bibnamefont {{Keating}}}, \bibinfo {author} {\bibfnamefont {V.}~\bibnamefont {{D'Odorico}}}, \bibinfo {author} {\bibfnamefont {R.~L.}\ \bibnamefont {{Davies}}}, \bibinfo {author} {\bibfnamefont {H.~M.}\ \bibnamefont {{Christenson}}}, \bibinfo {author} {\bibfnamefont {E.}~\bibnamefont {{Ba{\~n}ados}}}, \bibinfo {author} {\bibfnamefont {F.}~\bibnamefont {{Bian}}}, \bibinfo {author} {\bibfnamefont {M.}~\bibnamefont {{Bischetti}}}, \bibinfo {author} {\bibfnamefont {H.}~\bibnamefont {{Chen}}}, \bibinfo {author} {\bibfnamefont {F.~B.}\ \bibnamefont {{Davies}}}, \bibinfo {author} {\bibfnamefont {A.-C.}\ \bibnamefont {{Eilers}}}, \bibinfo {author} {\bibfnamefont {X.}~\bibnamefont {{Fan}}}, \bibinfo {author} {\bibfnamefont {P.}~\bibnamefont {{Gaikwad}}},
  \bibinfo {author} {\bibfnamefont {B.}~\bibnamefont {{Greig}}}, \bibinfo {author} {\bibfnamefont {M.~G.}\ \bibnamefont {{Haehnelt}}}, \bibinfo {author} {\bibfnamefont {G.}~\bibnamefont {{Kulkarni}}}, \bibinfo {author} {\bibfnamefont {S.}~\bibnamefont {{Lai}}}, \bibinfo {author} {\bibfnamefont {A.}~\bibnamefont {{Pallottini}}}, \bibinfo {author} {\bibfnamefont {Y.}~\bibnamefont {{Qin}}}, \bibinfo {author} {\bibfnamefont {E.~V.}\ \bibnamefont {{Ryan-Weber}}}, \bibinfo {author} {\bibfnamefont {F.}~\bibnamefont {{Walter}}}, \bibinfo {author} {\bibfnamefont {F.}~\bibnamefont {{Wang}}}, \ and\ \bibinfo {author} {\bibfnamefont {J.}~\bibnamefont {{Yang}}},\ }\href {\doibase 10.3847/1538-4357/ac6e60} {\bibfield  {journal} {\bibinfo  {journal} {\apj}\ }\textbf {\bibinfo {volume} {932}},\ \bibinfo {eid} {76} (\bibinfo {year} {2022})},\ \Eprint {http://arxiv.org/abs/2205.04569} {arXiv:2205.04569 [astro-ph.CO]} \BibitemShut {NoStop}%
\bibitem [{\citenamefont {{Zhu}}\ \emph {et~al.}(2023)\citenamefont {{Zhu}}, \citenamefont {{Becker}}, \citenamefont {{Christenson}}, \citenamefont {{D'Aloisio}}, \citenamefont {{Bosman}}, \citenamefont {{Bakx}}, \citenamefont {{D'Odorico}}, \citenamefont {{Bischetti}}, \citenamefont {{Cain}}, \citenamefont {{Davies}}, \citenamefont {{Davies}}, \citenamefont {{Eilers}}, \citenamefont {{Fan}}, \citenamefont {{Gaikwad}}, \citenamefont {{Haehnelt}}, \citenamefont {{Keating}}, \citenamefont {{Kulkarni}}, \citenamefont {{Lai}}, \citenamefont {{Ma}}, \citenamefont {{Mesinger}}, \citenamefont {{Qin}}, \citenamefont {{Satyavolu}}, \citenamefont {{Takeuchi}}, \citenamefont {{Umehata}},\ and\ \citenamefont {{Yang}}}]{zhu23}%
  \BibitemOpen
  \bibfield  {author} {\bibinfo {author} {\bibfnamefont {Y.}~\bibnamefont {{Zhu}}}, \bibinfo {author} {\bibfnamefont {G.~D.}\ \bibnamefont {{Becker}}}, \bibinfo {author} {\bibfnamefont {H.~M.}\ \bibnamefont {{Christenson}}}, \bibinfo {author} {\bibfnamefont {A.}~\bibnamefont {{D'Aloisio}}}, \bibinfo {author} {\bibfnamefont {S.~E.~I.}\ \bibnamefont {{Bosman}}}, \bibinfo {author} {\bibfnamefont {T.}~\bibnamefont {{Bakx}}}, \bibinfo {author} {\bibfnamefont {V.}~\bibnamefont {{D'Odorico}}}, \bibinfo {author} {\bibfnamefont {M.}~\bibnamefont {{Bischetti}}}, \bibinfo {author} {\bibfnamefont {C.}~\bibnamefont {{Cain}}}, \bibinfo {author} {\bibfnamefont {F.~B.}\ \bibnamefont {{Davies}}}, \bibinfo {author} {\bibfnamefont {R.~L.}\ \bibnamefont {{Davies}}}, \bibinfo {author} {\bibfnamefont {A.-C.}\ \bibnamefont {{Eilers}}}, \bibinfo {author} {\bibfnamefont {X.}~\bibnamefont {{Fan}}}, \bibinfo {author} {\bibfnamefont {P.}~\bibnamefont {{Gaikwad}}}, \bibinfo {author} {\bibfnamefont {M.~G.}\ \bibnamefont {{Haehnelt}}},
  \bibinfo {author} {\bibfnamefont {L.~C.}\ \bibnamefont {{Keating}}}, \bibinfo {author} {\bibfnamefont {G.}~\bibnamefont {{Kulkarni}}}, \bibinfo {author} {\bibfnamefont {S.}~\bibnamefont {{Lai}}}, \bibinfo {author} {\bibfnamefont {H.-X.}\ \bibnamefont {{Ma}}}, \bibinfo {author} {\bibfnamefont {A.}~\bibnamefont {{Mesinger}}}, \bibinfo {author} {\bibfnamefont {Y.}~\bibnamefont {{Qin}}}, \bibinfo {author} {\bibfnamefont {S.}~\bibnamefont {{Satyavolu}}}, \bibinfo {author} {\bibfnamefont {T.~T.}\ \bibnamefont {{Takeuchi}}}, \bibinfo {author} {\bibfnamefont {H.}~\bibnamefont {{Umehata}}}, \ and\ \bibinfo {author} {\bibfnamefont {J.}~\bibnamefont {{Yang}}},\ }\href {\doibase 10.3847/1538-4357/aceef4} {\bibfield  {journal} {\bibinfo  {journal} {\apj}\ }\textbf {\bibinfo {volume} {955}},\ \bibinfo {eid} {115} (\bibinfo {year} {2023})},\ \Eprint {http://arxiv.org/abs/2308.04614} {arXiv:2308.04614 [astro-ph.CO]} \BibitemShut {NoStop}%
\bibitem [{\citenamefont {{Fan}}\ \emph {et~al.}(2024)\citenamefont {{Fan}}, \citenamefont {{Chen}}, \citenamefont {{Avestruz}},\ and\ \citenamefont {{Khadir}}}]{fanMFPCROC2024}%
  \BibitemOpen
  \bibfield  {author} {\bibinfo {author} {\bibfnamefont {J.}~\bibnamefont {{Fan}}}, \bibinfo {author} {\bibfnamefont {H.}~\bibnamefont {{Chen}}}, \bibinfo {author} {\bibfnamefont {C.}~\bibnamefont {{Avestruz}}}, \ and\ \bibinfo {author} {\bibfnamefont {A.}~\bibnamefont {{Khadir}}},\ }\href {\doibase 10.48550/arXiv.2405.00100} {\bibfield  {journal} {\bibinfo  {journal} {arXiv e-prints}\ ,\ \bibinfo {eid} {arXiv:2405.00100}} (\bibinfo {year} {2024})},\ \Eprint {http://arxiv.org/abs/2405.00100} {arXiv:2405.00100 [astro-ph.CO]} \BibitemShut {NoStop}%
\end{thebibliography}%

\begin{figure*}[!h]
    \centering
    \includegraphics[width=0.5\textwidth]{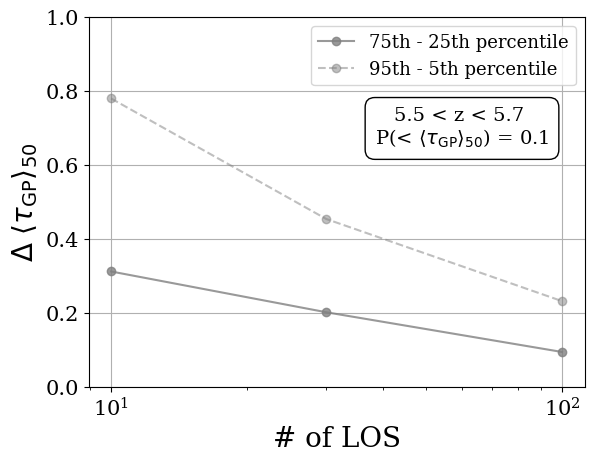}\hfill
    \includegraphics[width=0.5\textwidth]{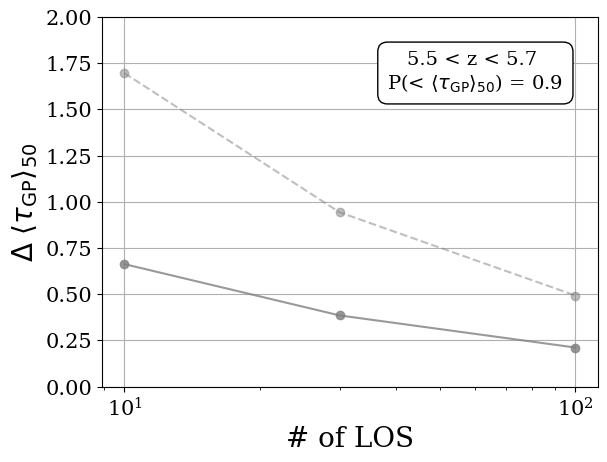}\hfill
    \includegraphics[width=0.5\textwidth]{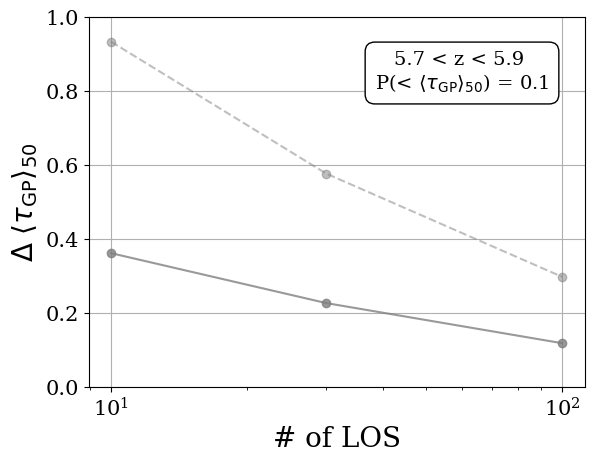}\hfill
    \includegraphics[width=0.5\textwidth]{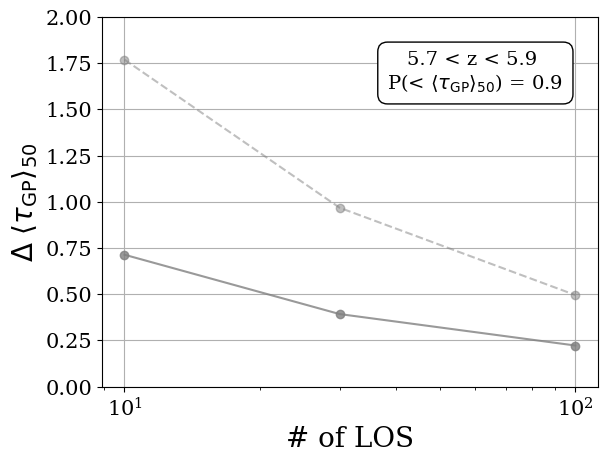}\hfill
    \includegraphics[width=0.5\textwidth]{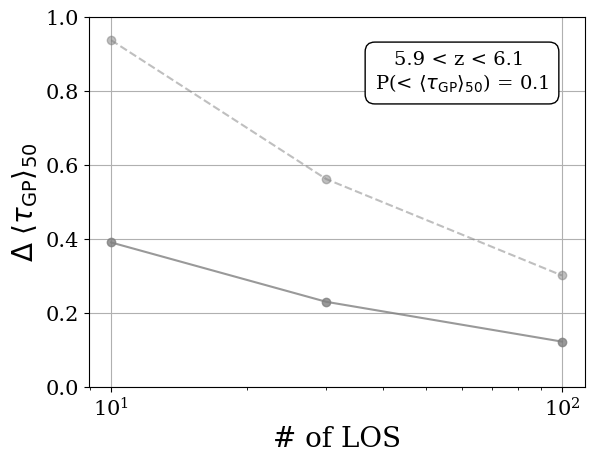}\hfill
    \includegraphics[width=0.5\textwidth]{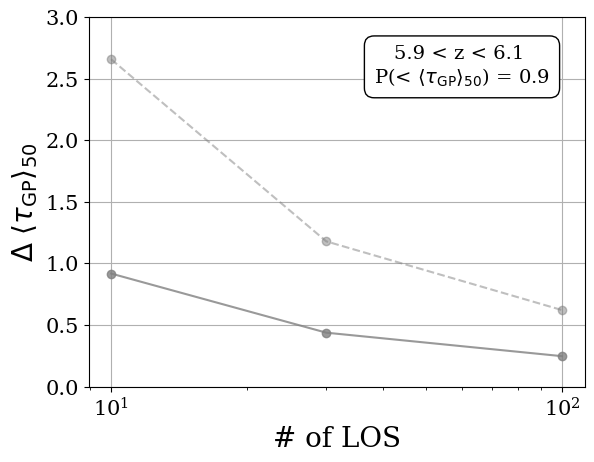}\hfill
    \caption{Width of the CDF at fixed values of $P(<\left\langle \tau_\mathrm{GP}\right\rangle_{50})=0.1$ (left panel) and $P(<\left\langle \tau_\mathrm{GP}\right\rangle_{50})=0.9$ (right panel)  as a function of number of sightlines. We show both the width of the 75th-25th percentile and the 95th-5th percentile.  The top, middle, and bottom rows correspond to the mean opacity distribution of CROC sightlines drawn from the $5.5<z<5.7$, $5.7<z<5.9$, and $5.9<z<6.1$ redshift ranges, respectively.}
    \label{fig:Varlos_appendix}
\end{figure*}

\begin{appendix}
In this Appendix, we provide a detailed examination of the CROC data, including figures that illustrate the variability of the distribution of mean opacities (without the overlaid comparisons to observational data, Fig.~\ref{fig:croc-pdf-tau} and Fig.~\ref{fig:croc-pdf-tau-30}). We also analyze how the width of the CDF evolves width the number of sightlines per subsample at other CDF values besides $0.5$ (Fig.~\ref{fig:Varlos_appendix}). 

\begin{figure*}[!h]
    \centering
    \includegraphics[width=0.5\textwidth]{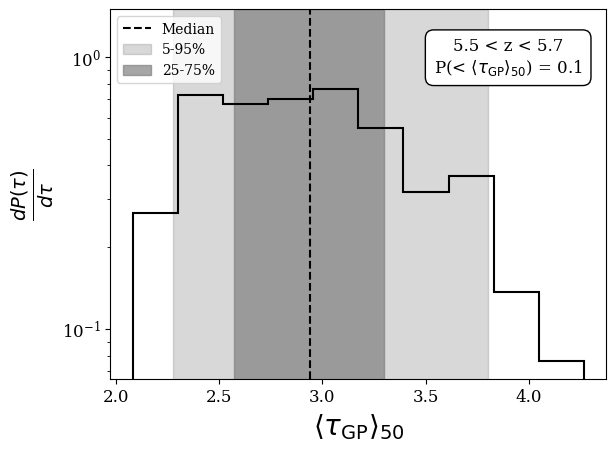}\hfill
    \includegraphics[width=0.5\textwidth]{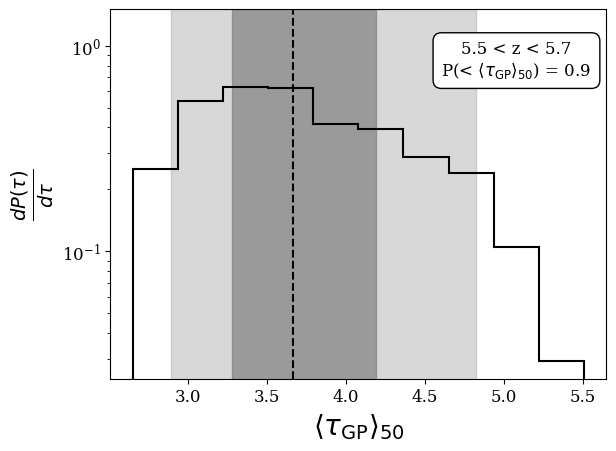}\hfill
    \caption{The distribution of mean opacities at the CDF values of $=0.1$ (left panel) and $=0.9$ (right panel) points from all subsamples of size 10 sightlines for the simulated sample representing the $5.7 < z < 5.9$ redshift bin. Similarly to the CDF $=0.5$ point in Fig.~\ref{fig:histogram}, the distribution is non-Gaussian. The median is indicated with a black dashed lines, with the dark gray band illustrating the 25th-75th percentile and the light gray band showing the 5th-95th percentile.} 
    \label{fig:appendix_histograms}
\end{figure*}

In Fig.~\ref{fig:appendix_histograms}, we show the distribution of mean opacities at fixed CDF values of 0.1 and 0.9 (direct analogs of Fig.~\ref{fig:histogram}). Specifically, these figures look at the probability density in the redshift range $5.5< z < 5.7$. Throughout all the redshift ranges and at all CDF values we consider the distributions are not Gaussian. We characterize the spread of these histograms with the 25th-75th percentile and 5th-95th percentile ranges (to capture the asymmetry and non-Gaussian tails of the distribution). 

\begin{figure*}[!h]
    \centering
    \includegraphics[width=0.5\textwidth]{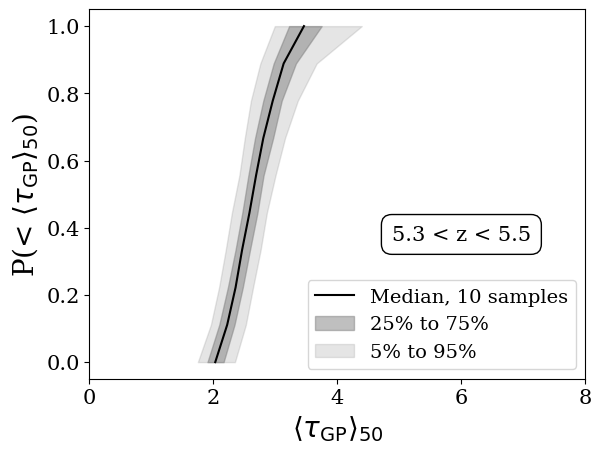}\hfill
    \includegraphics[width=0.5\textwidth]{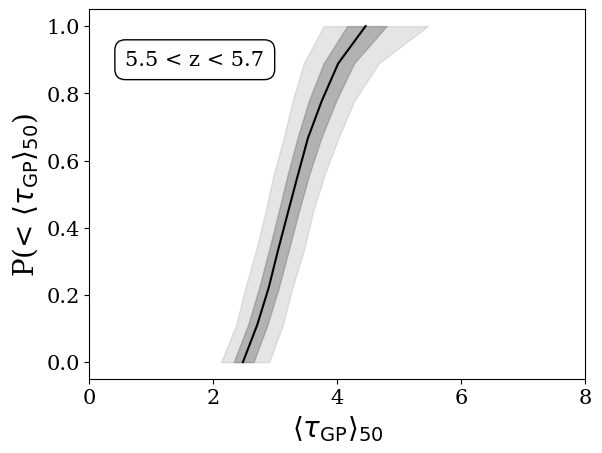}\\[1ex] 
    \includegraphics[width=0.5\textwidth]{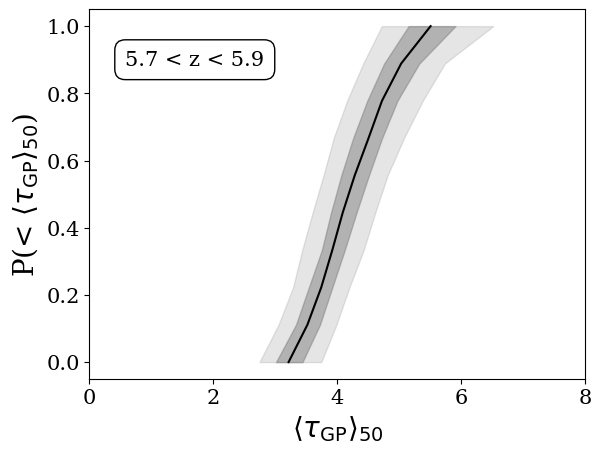}\hfill
    \includegraphics[width=0.5\textwidth]{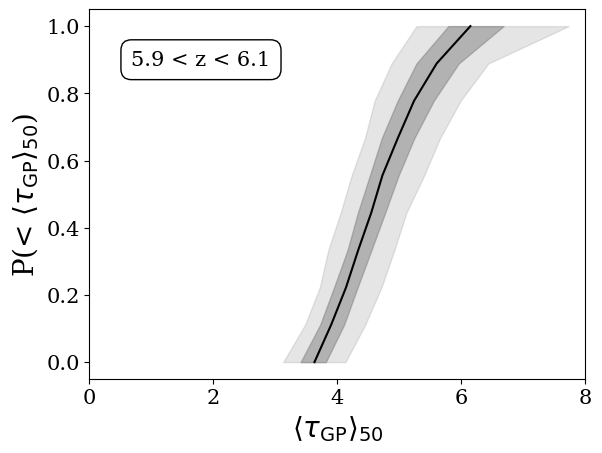}
    \caption{Cumulative probability distribution function (CDF) of mean opacity from CROC sightlines across four distinct redshift intervals. The CDFs are generated from subsamples of 10 randomly selected sightlines in all redshift bins. The shaded bands match those shown in Fig.~\ref{fig:appendix_histograms}. Among the four redshift ranges, the distribution corresponding to $5.7 < z < 5.9$ exhibits the broadest spread in mean opacities, indicating greater variability in absorption.}
    \label{fig:croc-pdf-tau}
\end{figure*}

Fig.~\ref{fig:croc-pdf-tau} and Fig.~\ref{fig:croc-pdf-tau-30} show the CDF of mean opacities from CROC simulations across four redshift bins, using a constant number of sightlines per subsample (10 in Fig.~\ref{fig:croc-pdf-tau} and 30 in Fig.~\ref{fig:croc-pdf-tau-30}). The CDFs shift to higher optical depth values with increasing redshift, where the neutral fraction of the IGM, and hence its optical depth to Ly$\alpha$ photons, is higher. As expected, the width of the CDF in each redshift range is smaller with 30 sightlines per subsample (Fig.~\ref{fig:croc-pdf-tau-30}) than with 10 sightlines per subsample (Fig.~\ref{fig:croc-pdf-tau}). Even using the same number of sightlines per subsample, the width of the distribution is larger in the higher redshift bins, with the tail at large $\langle \tau_\mathrm{GP} \rangle_{50}$ exhibiting the widest variance.

\begin{figure*}[!h]
    \centering
    \includegraphics[width=0.5\textwidth]{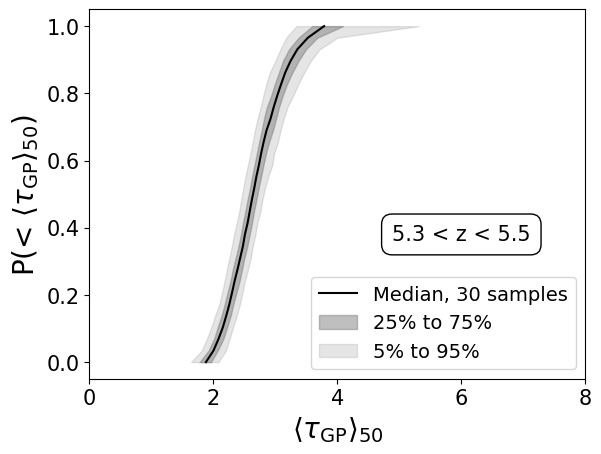}\hfill
    \includegraphics[width=0.5\textwidth]{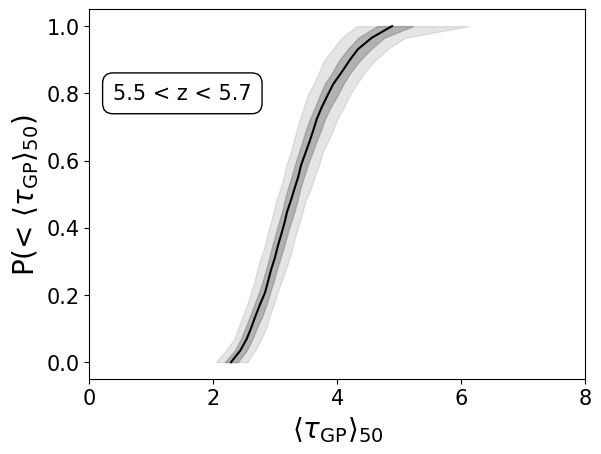}\\[1ex] 
    \includegraphics[width=0.5\textwidth]{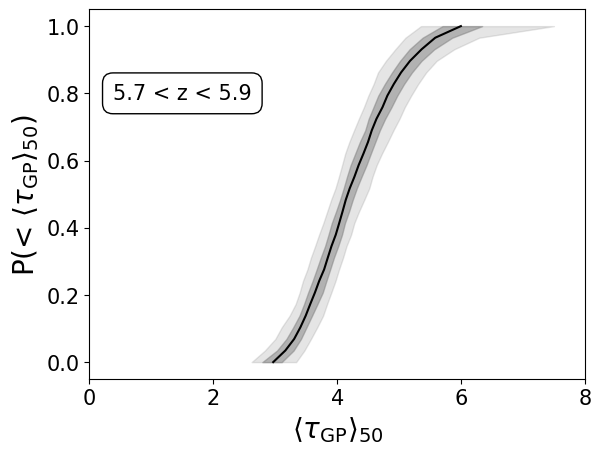}\hfill
    \includegraphics[width=0.5\textwidth]{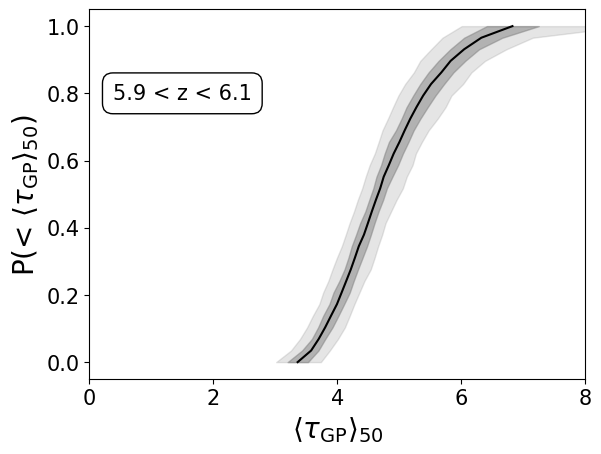}
    \caption{Cumulative probability distribution functions (CDFs) of mean opacities, with subsamples of 30 sightlines (similar to Fig.~\ref{fig:croc-pdf-tau}, but using larger subsamples here). Among the four redshift ranges, the distribution corresponding to $5.9 < z < 6.1$ exhibits the broadest spread in mean opacity values, indicating greater variability in absorption, and differing from the spread in Fig.~\ref{fig:croc-pdf-tau}.}
    \label{fig:croc-pdf-tau-30}
\end{figure*}

\begin{figure*}[!h]
    \centering
    \includegraphics[width=0.5\textwidth]{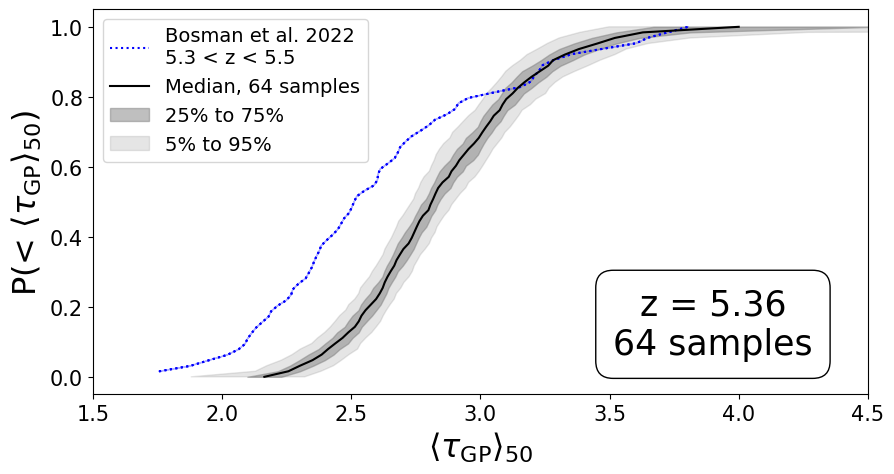}\hfill
    \includegraphics[width=0.5\textwidth]{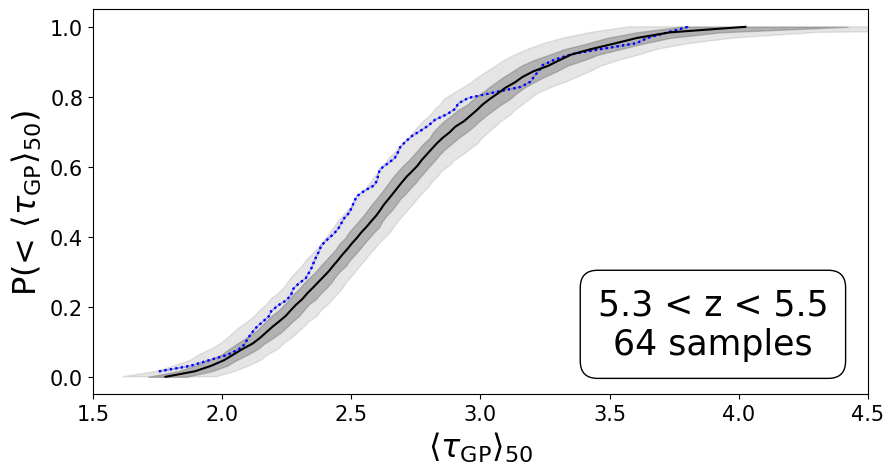}\\[1ex] 
    \includegraphics[width=0.5\textwidth]{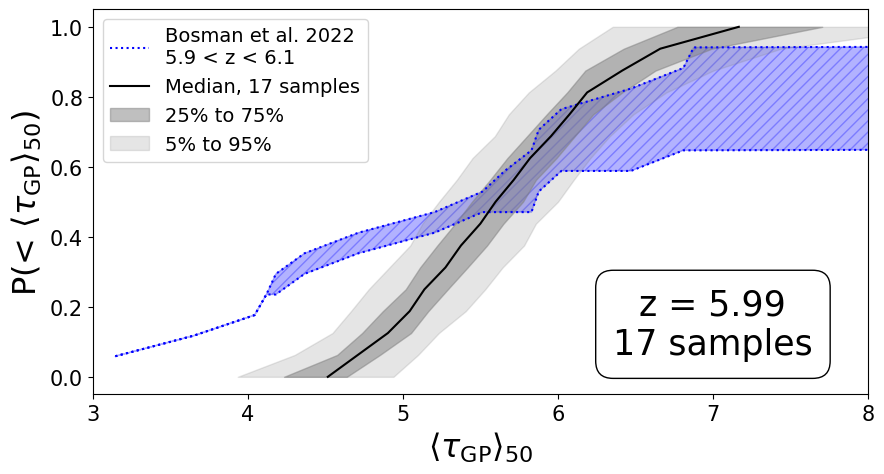}\hfill
    \includegraphics[width=0.5\textwidth]{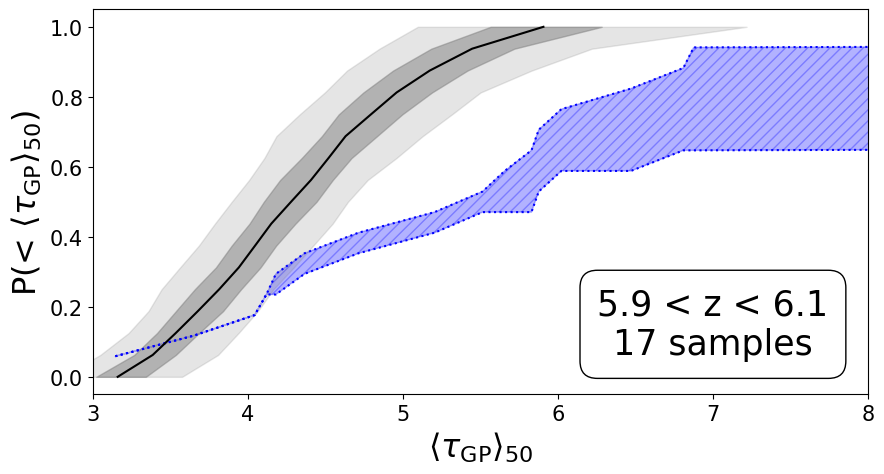}
    \caption{Cumulative probability distribution functions (CDFs) of mean opacities, with subsamples of matching size to observations.  Left panels show the CDF for sightlines drawn from a single snapshot at the midpoint of the redshift range, and right panels show the CDF for sightlines drawn from snapshots sampling the entire redshift range indicated. The mean opacity evolves sufficiently across the entire redshift range such that a comparison at the midpoint redshift leads to a bias where lower optical depths are undersampled.  When sampling from the entire redshift range instead of the midpoint snapshot, the CDF shifts to the left in both the highest and lowest redshift ranges. }
    \label{fig:croc-pdf-fixed-z-comparison}
\end{figure*}

In Fig.~\ref{fig:Varlos_appendix}, we show the width of the CDF for mean opacities from simulated sightlines at fixed CDF values of $P(< \langle \tau_\mathrm{GP} \rangle_{50}) = 0.1$ (left column) and 0.9 (right column), across three different redshift bins. This supplements Fig.~\ref{fig:Varlos}, which shows the CDF width at $P(< \langle \tau_\mathrm{GP} \rangle_{50}) = 0.5$ for the $5.5 < z< 5.7$ redshift range. Similar to Fig.~\ref{fig:Varlos}, both the 75th-25th percentile and 95th-5th percentile width decrease roughly as $N^{-1/2}$ (where $N$ is the number of sightlines per subsample) at $P(< \langle \tau_\mathrm{GP} \rangle_{50}) = 0.1$ and $0.9$. Fig.~\ref{fig:Varlos} also demonstrates that the width at $P(< \langle \tau_\mathrm{GP} \rangle_{50}) = 0.9$ is always larger than the corresponding width at $P(< \langle \tau_\mathrm{GP} \rangle_{50}) = 0.1$. 
\end{appendix}

\end{document}